\documentclass[11pt]{article}

\usepackage{fullpage}
\usepackage{mathptmx}
\usepackage[scaled]{helvet}
\usepackage{times}
\usepackage{latexsym}
\usepackage{amssymb,amsfonts,amsthm,amsmath}
\usepackage{graphicx}
\usepackage[dvipsnames,usenames]{color}
\usepackage{xspace}

\newcommand{\eat}[1]{}

\def\doctype{1}

\def\tsubmission{2}
\ifnum\doctype=\tsubmission
	\newcommand{\full}[1]{}
	\newcommand{\submit}[1]{#1}
\else
	\newcommand{\full}[1]{#1}
	\newcommand{\submit}[1]{}
\fi

\newtheorem{theorem}{Theorem}
\newtheorem{lemma}[theorem]{Lemma}

\newtheorem{claim}[theorem]{Claim}
\theoremstyle{definition}
\newtheorem{definition}{Definition}
\newtheorem{corollary}[theorem]{Corollary}

\newcommand{\opt}{\mathsf{opt}}

\newcommand{\algo}{\mathsf{algo}}
\newcommand{\pnorm}[2]{||#1||_{#2}}

\newcommand{\debmalya}[1]{{\color{red} Debmalya: #1}}

        {\hspace*{\fill}$\Box$\par}

\title{Online Load Balancing for Related Machines}
\author{Sungjin Im\footnote{Email: {\tt sim3@ucmerced.edu}. Supported in part by NSF grants CCF-1409130 and CCF-1617653.}\\UC Merced
\and Nathaniel Kell\footnote{Email: {\tt kell@cs.duke.edu}. Supported in part by NSF awards CCF-1527084 and CCF-1535972.}\\Duke University
\and Debmalya Panigrahi\footnote{Email: {\tt debmalya@cs.duke.edu}. Supported in part by NSF awards CCF-1527084 and CCF-1535972.}\\Duke University
\and Maryam Shadloo\footnote{Email: {\tt mshadloo@ucmerced.edu}. Supported in part by NSF grant CCF-1409130 and CCF-1617653.}\\UC Merced
}
\date{}

\begin{document}

\maketitle

\begin{abstract}

In the {\em load balancing} (or {\em job scheduling}) problem, introduced by Graham in the 1960s (SIAM J. of Appl. Math. 1966, 1969), 
jobs arriving online have to be assigned to machines so to minimize an objective defined on machine loads. A long
line of work has addressed this problem for both the makespan norm and arbitrary $\ell_q$-norms of machine loads. 
Recent literature 
(e.g., Azar {\em et al.}, STOC 2013; Im {\em et al.}, FOCS 2015) has further expanded the scope of this problem to 
{\em vector} loads, to capture jobs with multi-dimensional resource requirements in applications such as data centers. 
In this paper, we completely resolve the job scheduling 
problem for both scalar and vector jobs on {\em related} machines, i.e., where each machine has a given speed and 
the time taken to process a job is inversely proportional to the speed of the machine it is assigned on.
We show the following results:
\begin{itemize}
	\item {\em Scalar scheduling.}
    We give a constant competitive algorithm for optimizing any $\ell_q$-norm for (scalar) scheduling on related machines.
    The only previously known result was for the makespan norm.
    \item {\em Vector scheduling.}
    There are two natural variants for vector scheduling, depending on whether the speed of a machine is 
    dimension-dependent or not. We show a sharp contrast between these two variants,
    proving that they are respectively equivalent to unrelated machines and identical machines
    for the makespan norm. We also extend these results to arbitrary $\ell_q$-norms of the machine loads.
    No previous results were known for vector scheduling of related machines.
\end{itemize}

A key component of our algorithms is a new tool that we call {\em machine smoothing}, where we replace 
an arbitrary instance with a {\em smoothed instance} of the problem. The structural properties of the 
smoothed instance make it much simpler to argue about various norms of machine loads. We hope that this
generic technique will find more applications in other scheduling problems as well.

\end{abstract}

\thispagestyle{empty} \setcounter{page}{0} \clearpage

\section{Introduction}
\label{sec:introduction}

The {\em load balancing} (or {\em job scheduling}) problem, introduced in the seminal work of Graham in the 1960s~\cite{Graham66,Graham69},
asks for an online assignment of jobs to machines so as to minimize some objective defined on machine loads. 
A long line of work has addressed this problem for both the makespan norm (maximum load) and for other 
$\ell_q$-norms of machine loads (e.g.,~\cite{BartalFKV95,KargerPT96,Albers99,FleischerW00,Albers99, FaigleKT89, BartalKR94, GormleyRTW00, Rudin01,AspnesAFPW97,AzarNR95,BermanCK00,AvidorAS01,AwerbuchAGKKV95,Caragiannis08}). In this 
paper, we study this problem in the {\em related} machines setting, where the processing time of a job on a machine is 
inversely proportional to the speed of the machine. The only previous result for this problem on related machines 
was a constant-competitive algorithm for the 
makespan (maximum load) objective~\cite{BermanCK00}. However, in many situations, other $\ell_q$-norms of machine loads are more relevant: e.g., 
the $2$-norm is suitable for disk storage~\cite{ChandraW75,CodyC76}, whereas $q$ between 2 and 3 is used for modeling 
energy consumption~\cite{Pelley09,Albers10,YaoDS95}. This led to constant-competitive algorithms for arbitrary 
$\ell_q$-norms of machine loads for the special case of identical machines (all machine speeds are equal)~\cite{AvidorAS01},
and to $O(q)$-competitive algorithms for the more general unrelated machines setting (processing times are 
arbitrary)~\cite{AwerbuchAGKKV95,Caragiannis08}. But, this problem has remained open for related machines. 

Moreover, recent literature has further expanded the scope of the job scheduling problem 
to vector jobs that have multiple dimensions, the resulting problem being called 
{\em vector scheduling}~\cite{ChekuriK04,AzarCKS13,MeyersonRT13,ImKM15}. This problem is very relevant to
scheduling on data centers where jobs with multiple resource requirements have to be allocated to machine 
clusters to make efficient use of limited resources such as CPU, memory, network bandwidth, and 
storage~\cite{drf,popa2012faircloud,lee2011heterogeneity,ColeGG,ImKM14,ImKM15}. Recently, Im~{\em et al.} \cite{ImKKP15} showed that for vector scheduling with the makespan norm, competitive ratios of 
$O(\log d / \log \log d)$ and $O(\log d + \log m)$ are tight for identical and unrelated machines respectively,
where $d$ is the number of dimensions and $m$ is the number of machines. They also extended these results
to arbitrary $\ell_q$-norms.
In many data center applications, the situation is between these two extremes of identical and unrelated machines, 
and resembles the related machines scenario. In other words, machines have non-uniform speeds and the load created 
a vector job on any dimension of a machine is inversely proportional to the machine speed. But,
vector scheduling for related machines had not been addressed previously, either for the makespan norm 
or for arbitrary $\ell_q$ norms.

\smallskip
We completely resolve these two sets of problems for scalar and vector scheduling on related machines in this paper.
Our first result is for the scalar setting, and gives {\em a constant-competitive algorithm 
for optimizing any $\ell_q$-norm of machine loads on related machines.} 
In previous work, the constant competitive ratio for makespan on related machines was obtained by the so-called {\em slowest-fit} 
algorithm~\cite{BermanCK00}. The main idea in this algorithm is to guess the optimal makespan, and assign a job arriving
online to the slowest machine that can accommodate it without exceeding the optimal makespan by a constant factor. 
But, this strategy fails for other $\ell_q$-norms. Even if we were to guess the optimal value of the norm,
this does not tell us the relative contributions of the different machines to the optimal objective. 
Therefore, guessing the optimal value is not 
sufficient to fix bounds on the loads of individual machines  (unlike makespan,
where the guessed optimum gives a bound for the load on each machine). This rules out 
an assignment strategy like slowest-fit. Instead, we develop a new tool that we call {\em machine smoothing},
and use it in all our algorithms. Before describing this idea, let us turn to
vector scheduling and describe our results for this problem.


\smallskip
 Our next contribution in this paper is to \emph{resolve the online vector scheduling problem for related machines}. 
 We show that if machine speeds are dimension-independent (we call this the {\em homogeneous} case), then the competitive ratio 
 asymptotically matches that of identical machines for the makespan norm. We also extend this result to arbitrary $\ell_q$-norms. 
 On the other hand, we show that if machine speeds are dimension-dependent (we call this the {\em heterogeneous} case), then the 
 competitive ratio asymptotically matches that of unrelated machines. Both homogeneous and heterogeneous speeds are relevant 
 to the practical context and respectively represent situations where clusters only differ in the number of machines or in 
 machine types as well.\footnote{Note that by scaling, it is sufficient in the homogeneous case for the speeds on different 
 resources to be proportional -- they do not need to be exactly equal.} 
 Unfortunately, the slowest-fit algorithm does not work for vector scheduling on homogeneous machines, 
 even for the makespan norm (see Appendix~\ref{app:example} for a counterexample). As with scalar scheduling, we again resort to the 
 machine smoothing idea that we describe next.

\smallskip
From a technical perspective, 
a key tool in our algorithms is what we call {\em machine smoothing}. Imagine grouping together machines with similar speeds.
Then, one can employ a two-stage algorithm that assigns each job to a machine group, and then 
employs an identical machines algorithm within each machine group. But, how do we figure out an assignment of 
jobs to machine groups? The number of machines in each group might be completely arbitrary,
making such assignment a challenging problem. It turns out that the assignment
of jobs to groups is facilitated if we can ensure that the cumulative {\em processing power} in a group 
exponentially increases as we move to slower groups.
(The cumulative processing power for the makespan objective is simply the sum of speeds of machines in the group;
for other $\ell_q$-norms, this definition is suitably generalized.)
So, now we have two objectives: group machines with similar speeds,
but also ensure exponentially increasing processing powers of the groups in decreasing speed order. 
To simultaneously satisfy these
goals, we define a machine smoothing procedure that initially groups machines to satisfy the second condition,
but then replaces the machines of non-uniform speeds in a group by a suitably defined {\em equivalent} set
of identical machines. We show that this generic transformation can be performed for any given instance, and
for any $\ell_q$-norm, while only sacrificing a constant factor in the competitive ratio of the algorithm.
We call this transformed instance a {\em smoothed instance} of the problem.

It turns out that the machine smoothing technique is essentially sufficient for solving the makespan 
minimization problem in vector scheduling,
since the assignment of jobs to machine groups in a smoothed instance can be done by simulating the 
slowest-fit strategy used for scalar scheduling. However, for other $\ell_q$-norms, even for scalar 
scheduling, we need to work harder
in designing the algorithm to assign jobs to machine groups in a smoothed instance. In particular, we use a two-step
approach. First, we use a gradient descent algorithm on a suitably chosen fractional relaxation of the norm
to produce a competitive fractional solution. Next, we use an online rounding algorithm to produce an integer
assignment from the fractional solution. In the case of vector scheduling for arbitrary $\ell_q$-norms,
an additional complication is caused by the fact that the gradient descent algorithm can produce unbalanced 
loads on different dimensions since it follows the gradient for a single objective, thereby leading to a 
large competitive ratio. To avoid this difficulty, we use the assignment produced by the gradient 
descent algorithm only as an {\em advice} on the approximate speed of the machine group that a fractional job should be assigned
to. We then use a different algorithm to make the actual assignment of the fractional job to a machine group similar
to the advice, but not necessarily to the exact same group. Interestingly, while 
identical machines admit algorithms that optimize all norms 
{\em simultaneously}~\cite{ImKKP15}, we rule this out for homogeneous related machines (Appendix~\ref{sec:no-all-norms}). 
Therefore, our algorithms for vector 
scheduling for arbitrary $\ell_q$-norms use the value of $q$ in the algorithm itself, and this is necessary given our 
lower bound on optimizing all norms simultaneously.

For the heterogeneous setting, a simple adaptation of the unrelated machines lower bound of $\Omega(\log m)$ 
gives an instance with $d = \Omega(m)$. This is not interesting because a dependence on $\log d$ is required even for identical machines. Instead, we design an encoding scheme that uses only $d = O(\log m)$ but still manages to show a lower bound of $\Omega(\log m)$.
The makespan lower bound for heterogeneous related machines extends to other norms as well, thereby matching known bounds for unrelated machines 
for all $\ell_q$-norms.

\medskip
\noindent
{\bf Preliminaries and Results:}
First, we set up some standard notation.
In online scheduling, a set of $n$ jobs arrive online and each job must be irrevocably assigned to one of $m$ machines immediately on arrival. Each job $j$ has a non-negative \emph{size} $p_j$. In vector scheduling, $p_j$ is a vector of $d$ dimensions, $p_j = \langle p_j(1), p_j(2), \ldots, p_j(d) \rangle$. Each machine $i$ has a non-negative speed $s_i$ that is given offline. In vector scheduling, $s_i$ is a vector $\langle s_i(1), s_i(2), \ldots, s_i(d) \rangle$, where $s_i(1) = s_i(2) = \ldots = s_i(d)$ (denoted $s_i$) in the homogeneous setting. 
When job $j$ is assigned to machine $i$, it produces a {\em load} of $p_j/s_i$. In vector scheduling, the load is $p_j(k) / s_i(k) = p_{ij}(k)$ in dimension $k$. The load produced by a set of jobs is the sum of their individual loads. The {\em load vector} is denoted $\Lambda = \langle \Lambda_1, \Lambda_2, \ldots, \Lambda_m \rangle$, where $\Lambda_i$ is the total load on machine $i$. For vector scheduling, every dimension $k$ has its own load vector, denoted $\Lambda(k) = \langle \Lambda_1(k), \Lambda_2(k), \ldots, \Lambda_m(k) \rangle$, where $\Lambda_i(k)$ is the total load on machine $i$ in dimension $k$.

In vector scheduling, the makespan objective is given by:
\vspace{-1.5mm}
\begin{equation*}
	\max_{k=1}^d ||\Lambda(k)||_{\infty} = \max_{k=1}^d \max_{i=1}^m \Lambda_i(k).
\end{equation*}
\vspace{-1.5mm}
For the problem of minimizing makespan in vector scheduling, we show the following result.
\begin{theorem}
\label{thm:related-makespan}
	For online vector scheduling on related machines for minimizing makespan:
	\vspace{-1.5mm}
	\begin{enumerate}
		\setlength{\itemsep}{0pt}
	\setlength{\parskip}{0pt}
	\item (Section~\ref{sec:homogeneous-related}) For homogeneous speeds, we give a deterministic algorithm with a competitive ratio of $O(\log d / \log \log d)$. This is asymptotically tight since it matches a known lower bound for identical machines~\cite{ImKKP15}. 
	\item (Section~\ref{sec:hetero-related})   For heterogeneous speeds, we give a lower bound of $\Omega(\log d + \log m)$ on the competitive ratio. This is asymptotically tight since it matches a known upper bound for unrelated machines~\cite{MeyersonRT13,AzarCKS13,ImKKP15}.
	\end{enumerate}
\end{theorem}
\vspace{-1mm}

Now we state our results for optimizing arbitrary $\ell_q$-norms. First, we consider the scalar
scheduling problem. The $\ell_q$-norm objective is given by (we often call this just the $q$-norm, for brevity):
\begin{equation*} 
	|| \Lambda||_{q} = \left(\sum_{i = 1}^m (\Lambda_i)^q\right)^{1/q}
\end{equation*}
We obtain the following result.
\begin{theorem}
\label{thm:scalar}
	For online (scalar) scheduling on related machine for minimizing $\ell_q$-norms:
    \vspace{-1.5mm}
	\begin{enumerate}
		\setlength{\itemsep}{0pt}
	\setlength{\parskip}{0pt}
	\item (Section~\ref{sec:scalar-q} and Section~\ref{app:scalar-q}) We give a deterministic algorithm with a constant competitive ratio. This is asymptotically tight because online scheduling has a constant lower bound even for identical machines~\cite{Albers99, FaigleKT89, BartalKR94, GormleyRTW00, Rudin01}. 
	\end{enumerate}
\end{theorem}

Next, we consider optimizing $\ell_q$-norms in vector scheduling.
our objective is given by:
\begin{equation*} 
\max_{k=1}^d || \Lambda(k)||_{q} = \max_{k=1}^d \left(\sum_{i = 1}^m (\Lambda_i(k))^q\right)^{1/q}
\end{equation*}
We obtain the following result.
\begin{theorem}
\label{thm:related-pnorm}
	For online vector scheduling on related machines for minimizing $\ell_q$-norms:
	\vspace{-1.5mm}
	\begin{enumerate}
		\setlength{\itemsep}{0pt}
	\setlength{\parskip}{0pt}
	\item (Section~\ref{sec:vector-q} and Section~\ref{app:vector-q}) For homogeneous speeds, we give a deterministic algorithm with a competitive ratio of $O(\log^c d)$ for some constant $c$. This is tight up to the value of the constant 
    $c$, by a known lower bound for identical machines~\cite{ImKKP15}. 
	\item (Section~\ref{sec:hetero-related})   For heterogeneous speeds, we give a lower bound of $\Omega(\log d + q)$ on the competitive ratio. This is asymptotically tight since it matches a known upper bound for unrelated machines~\cite{ImKKP15}.
	\end{enumerate}
\end{theorem}

Note that Theorem~\ref{thm:scalar} follows as a corollary of Theorem~\ref{thm:related-pnorm}.
However, our vector scheduling algorithm uses our scalar scheduling algorithm as a subroutine;
consequently, the proof of Theorem~\ref{thm:related-pnorm} relies on an independent proof
of Theorem~\ref{thm:scalar}. Therefore, we present our scalar scheduling results before presenting
our vector scheduling results for arbitrary $q$-norms.

\smallskip
\noindent
{\bf Related Work.}
In the interest of space, we will only state a small subset of related results and refer the reader to more detailed surveys
~\cite{Azar96, Sgall96, PruhsST04} for other results.

The online job scheduling problem was introduced by Graham~\cite{Graham66}, who showed that list scheduling has a competitive ratio of $(2-1/m)$ for the makespan objective on identical machines. Currently, the best known upper bound is $1.9201$ \cite{BartalFKV95,KargerPT96,Albers99,FleischerW00}, while the best lowerbound is $1.880$ \cite{Albers99, FaigleKT89, BartalKR94, GormleyRTW00, Rudin01}. For the related machines setting, the slowest-fit algorithm is 2-competitive \cite{BermanCK00}, but for  unrelated machines, the optimal competitive ratio is $\Theta(\log m)$ \cite{AzarNR95, AspnesAFPW97}.
  This problem was generalized to arbitrary $q$-norms by \cite{AvidorAS01} for identical machines and \cite{AwerbuchAGKKV95,Caragiannis08} for unrelated machines. The only previous result for related machines was the competitive ratio of $2$ achieved by the slowest-fit
algorithm for the makespan norm~\cite{BermanCK00}.

The multidimensional version of this problem was introduced by Chekuri and Khanna in the offline model~\cite{ChekuriK04}, who gave a PTAS for constant $d$. For unrelated machines, they showed a constant lower bound, and the best known approximation factor is $O(\log d/\log \log d)$ due to Harris and Srinivasan~\cite{HarrisS13}. In the online setting, Azar \emph{et al.} \cite{AzarCFR16} and Meyerson \emph{et al.} \cite{MeyersonRT13} gave $O(\log d)$-competitive algorithms for identical machines. Recently, Im \emph{et al.} \cite{ImKKP15} improved these results by giving tight bounds of $O(\log d/\log \log d)$ for identical machines and $O(\log d + \log m)$ for unrelated machines. They also extended these results to arbitrary $q$-norms, giving tight bounds of $O((\frac{\log d}{\log \log d})^{\frac{q-1}{q}})$ and 
$O(\log d + q)$ for identical and unrelated machines.

\smallskip
\noindent
{\bf Roadmap.} In the next section, we present the idea of machine smoothing that is a generic tool 
we use in all the algorithms. This is essentially sufficient for minimizing makespan in vector scheduling
on homogeneous machines (Section \ref{sec:homogeneous-related}), but we need more 
ideas for minimizing arbitrary $q$-norms. Most of these new ideas are for the fractional algorithms,
which we present in Sections~\ref{sec:scalar-q} and \ref{sec:vector-q} for scalar and vector
scheduling respectively.
The corresponding rounding algorithms are presented in Sections~\ref{app:scalar-q} and
\ref{app:vector-q}, respectively. Finally, in 
Section~\ref{sec:hetero-related}, we present our lower bounds for vector scheduling on
heterogeneous machines. 

\section{Machine Smoothing}
\label{sec:grouping}

One of the main ideas that we use throughout our algorithms is that of {\em machine smoothing}.
There are two properties that we wish to derive from machine smoothing: that machines
in a single group have the same speed and that a slower group has processing power 
at least as much the sum over all its faster groups. To ensure both properties simultaneously,
simply grouping the given machines is not sufficient -- instead, we need to modify machine 
speeds in the given instance. The goal of this section is to show that such modification
is valid, i.e., it does not significantly change the optimal objective.

We will describe the machine smoothing procedure for an arbitrary $q$-norm objective.
First, we articulate the properties that we demand at the end of the transformation.
\begin{definition}
	\label{def:canonical-gen}
	We say that machines in an instance are {\em smoothed }if they can be partitioned into groups, $G_0, G_1, G_2, G_3, \cdots $ such that: \vspace{-1.5mm}
\begin{itemize}
	\setlength{\itemsep}{0pt}
	\setlength{\parskip}{0pt}
	\item  \textbf{Property 1}: All machines in each group have \emph{equal} speed.
	\item  \textbf{Property 2}: $S(G_l) := \sum_{i \in G_l} s_i^{\gamma} \geq S(G_0) + S(G_1) + ... + S(G_{l-1})$, where $\gamma = q / (q-1)$.
	\item  \textbf{Property 3}: For any two groups $G_l$ and $G_{l'}$ where $l < l'$, any machine in group $G_l$ has a higher speed  than any machine in $G_{l'}$ -- if two machines have different speeds, their speed differ by at least a factor of 2. 
 \end{itemize}
 \end{definition}
 
The next lemma claims that any instance can be transformed into a smoothed instance without significantly
changing the optimal objective.

\begin{lemma}
	\label{lem:canonical-gen}
	For any set $M$ of machines (with homogeneous speeds in the case of vector scheduling), we can construct a smoothed set $M'$ of machines 
    such that for any set $J$ of jobs, the respective optimal solutions are related as $\opt(J, M') \leq O(1) \cdot \opt(J, M)$. Furthermore, 
    there exists a mapping $g: M' \rightarrow M$ such that if a job scheduled on a machine $i' \in M'$ is scheduled on machine $g(i') \in M$, 
    then the resulting $q$-norm for the original set $M$ of machines is at most a constant factor larger than the $q$-norm for the new set 
    $M'$ of machines.
\end{lemma}
\begin{proof} 
	%
  We assume (wlog, by scaling) that the fastest machine in $M$ has speed exactly 1. We also round all machine speeds to (negative) powers of $\Gamma := 2^{1/\gamma}$. We order machines in non-increasing order of their speeds, breaking ties arbitrarily. The first group $G_0$ is the singleton set that has only one machine with speed $1$. We now create the remaining groups inductively until every machine is assigned to a group. For $l \geq 1$, exclude machines in $G_0 \cup G_1 \cup ... \cup G_{l-1}$ and define $G_l$ to be the minimal set of the fastest machines $i$, whose sum of $s_i^{\gamma}$ is exactly $2^l$. This is always possible to do since we rounded the machine speeds to (negative) powers of $\Gamma$, hence $s_i^{\gamma}$ are (negative) powers of $2$. (The last group $G_{L+1}$ may not satisfy this property.) 
  
  Define $S(G) := \sum_{i \in G} s_i^\gamma$ for any group $G$. 	
	For each group $G_l$, note that $S(G_l) = 2^l$. Let $s_{\min}(G_l)$ denote the lowest speed of all machines in $G_l$. We replace $G_l$ with a new set $G'_l$ of machines whose speeds are all equal to $s_{\min}(G_l)$, such that $S(G'_l) = 2^l$. Let $M'$ denote the machines that we have constructed.


We now prove the first claim that the optimal $q$-norm increases by at most a constant factor for the new machines $M'$. Fix an optimal schedule. Since the first group doesn't change, i.e., $G_0 = G'_0$, any job assigned to the machine in $G_0$ stays there. If the optimal schedule assigns a job $j$ to a machine in $G_{l+1}$, $1 \leq l \leq L$, we move the job to a machine in $G'_l$.
We let each machine $i' \in G'_l$ process jobs assigned to $T := 2\cdot \frac{s'^\gamma}{s^\gamma}$ machines $i\in G_{l+1}$, where $s := s_i$ and $s' := s_{i'}$. Note that this is possible since $S(G'_l) = |G'_l|\cdot s'^\gamma = 2^l$ and $S(G_{l+1}) = |G_{l+1}|\cdot s^\gamma = 2^{l+1}$, which implies that $|G'_l|/|G_{l+1}| = 2\cdot \frac{s'^\gamma}{s^\gamma}$.  To see that the $q$-norm increases by a constant factor, consider a fixed dimension and let $u_1, u_2, ... , u_T$ be the volume of jobs assigned to $T$ machines on the fixed dimension. Then, we have
\begin{equation}
\label{eq:sim}
\sum_{t = 1}^T \left(\frac{u_t}{ s}\right)^q 
\geq T \cdot \left(\frac{\sum_{t = 1}^T u_t }{sT}\right)^q 
= \left(\frac{1}{T}\right)^{q-1} \cdot \left(\frac{s'}{s}\right)^q \cdot \left(\frac{\sum_{t = 1}^T u_t }{s'}\right)^q 
\geq \frac{1}{2^q} \left(\frac{\sum_{t = 1}^T u_t }{s'}\right)^q.
\end{equation}
This implies that the $q^q$-norm increases by a factor of at most $2^q$. The first group $G'_0$ processes jobs relocated not only from $G_1$ but also from $G_0$. Hence the $q^q$-norm increases by a factor of at most $4^q$, meaning that the optimal $q$-norm increases by a constant factor.

It now remains to prove the second claim. Consider any online algorithm $A$. If $A$ assigns a job to a machine $i' \in G'_l$, we assign it to a machine $i$ in $G_l$; we do not use any machine in $G_{L+1}$. Fix a group $G'_l$. We associate each machine with speed $s$ in $G_l$ with $T' := \frac{s^\gamma}{s'^\gamma}$ unique machines in $G_{l'}$ (all these machines have speed $s'$). This is possible since $S(G_l) = S(G'_l)$. Now, using a calculation identical to Eq.~\eqref{eq:sim}, we can conclude that the $q$-norm increases by at most a constant factor in this reassignment.

Also, note that the initial rounding of speeds is only by a constant factor, and hence this also changes the $q$-norm only by a constant factor. 
As a consequence, we can now claim that the two properties of the lemma are satisfied by the transformed set of machines $M'$.

Finally, we are left to prove that the set of machines $M'$ comprise a smoothed instance. 
It is straightforward to see that these machines, grouped in
$G'_0, G'_1, G'_2, \cdots, G'_L$, satisfy the first two properties of smoothed instances. For the third property, we first merge all groups
with the same speed. This does not affect the first two properties, and satisfies a weaker version of {\bf Property 3} where machine
speeds differ by at least a factor of $2^{1/\gamma}$. To improve this separation to a factor of $2$, we merge groups with
speeds $s'$ satisfying $2^l \leq s' < 2^{l+1}$ for each (non-positive) value of $i$. We now satisfy {\bf Property 2} and {\bf 3},
but not {\bf Property 1}. To satisfy {\bf Property 1} as well, we replace the machines of a group $G'_l$ with
speeds $2^l \leq s' < 2^{l+1}$ by a new group $G''_l$ containing machines of speed $2^l$ such that 
$\sum_{i'\in G'_l} s_{i'}^\gamma =  |G''_l|\cdot (2^l)^\gamma$. By mapping machines exactly as above (we omit details for brevity), 
we can bound the change in
the $q$-norm for both the algorithm and an optimal solution by a constant factor. It is easy to verify that the 
set of machine groups defined by $G''$ satisfy all the properties of a smoothed instance.
\end{proof}

We say that a group is lower than the other group if machines in the group have a lower speed.
Note that the set of machines is given to the algorithm a priori. Hence we can find $M'$ and the mapping $g$ offline, and using the mapping $g$ from $M'$ to $M$, we can convert an online algorithm for the smoothed instance into an online algorithm for the original instance. For this reason, we can assume wlog that machines are smoothed. Also, note that for the makespan norm, the above grouping works exactly as described by setting 
$\gamma = 1$. 
\vspace{-2.5mm}
\section{Vector Scheduling: Minimizing Makespan}
	\label{sec:homogeneous-related}
\vspace{-1.5mm}


In this section, we give our $O\left( \frac{\log d }{\log \log d}\right)$-competitive algorithm for makespan minimization on homogeneous related machines 
(the first part of Theorem~\ref{thm:related-makespan}).
Recall that in this setting,  machine $i$ has a uniform speed vector $\langle s_i, s_i, \cdots, s_i \rangle$, where we refer to $s_i$ as machine $i$'s speed. By scaling, we assume w.l.o.g that the highest speed of any machine is exactly 1. We assume throughout that we have a smoothed instance, 
which is wlog by Lemma~\ref{lem:canonical-gen}.

\vspace{-3mm}
\paragraph{Algorithm.} Since all machines in the same group have equal speed, we use $s_l$ to denote the speed of any machine in group $G_l$. For simplicity, we say that group $G_l$'s speed is $s_l$. We assume wlog that we know the value of the optimal makepsan, $\opt$ within a constant factor by using a standard doubling technique. We say that a group $G_l$ is permissible for job $j$ if $\max_k \frac{p_j(k)}{s_l} \leq \opt$. The algorithm has two components:
\begin{itemize}
	\setlength{\itemsep}{0pt}
	\setlength{\parskip}{0pt}
	\item Assigning jobs to groups of machines: Assign job $j$ to a permissible group $G_l$ with the largest index $l$; note that $G_l$ has the lowest speed among all permissible groups for job $j$. Let $J_l$ denote jobs assigned to group $G_l$. 
	\item Assigning jobs to machines within each group: For each group $G_l$, run the deterministic $O(\log d / \log \log d)$-competitive algorithm for identical machines in \cite{ImKKP15} for minimizing makespan to schedule jobs in $J_l$ on machines in $G_l$. 
\end{itemize}

We formally state the lower bound used in the analysis of the algorithm in \cite{ImKKP15} used above.

\begin{theorem}[\cite{ImKKP15}]
	\label{thm:focs-identical}
	Suppose that jobs arrive to be scheduled on $m$ identical machines. 
	For any $T$ such that $\max_{k,j} p_j(k) \leq T$ and $\max_{k} \sum_j p_j(k) / m \leq T$, then there is a 
	deterministic algorithm that yields a schedule with makespan $O\left(\frac{\log d}{\log \log d}\right) \cdot T$.
\end{theorem}

The competitive ratio of the algorithm is derived based on two obvious lower bounds, the maximum job size over all dimensions and the average load vectors over $m$ machines. We note that the theorem is stated under the assumption that $T$ is known to the algorithm a priori, but we can again easily remove this assumption by using a standard doubling technique. 

\smallskip
   We are now ready to complete the proof. Consider any fixed $l$. Since we schedule jobs $J_l$ on identical machines in $G_l$, it suffices to show that  $\max_{k,j \in J_l} \frac{p_j(k)}{s_l}  \leq O(1) \cdot \opt$ and $\max_{k} \frac{\sum_{j \in J_l} p_j(k)} {S(G_l)} \leq O(1) \cdot \opt$.
%
%
     Note that group $G_l$ is permissible for any job in $J_l$. Hence we have $\max_{k, j\in J_l} \frac{p_j(k)}{s_l} \leq \opt$. Since the optimal scheduler can schedule jobs in $J_l$ only on machines in $G_1 \cup G_2 \cup ... \cup G_l$ (i.e., $G_l$ is the slowest permissible group for jobs in $J_l$), we have for any dimension $k$, 
    $$\sum_{j \in J_l} p_j(k) 
    \leq \sum_{l'=0}^l S(G_{l'}) \cdot \opt 
    = (S(G_l) + \sum_{l'=0}^{l-1} S(G_{l'})) \cdot \opt 
    \leq S(G_l)\cdot (2\cdot \opt) \quad \text{(by {\bf Property 2} of smoothed instances)}.$$
 Thus, by Theorem~\ref{thm:focs-identical}, the makespan of machines $G_l$ is  $O\left(\frac{\log d}{\log \log d}\right) \cdot  \opt$. 
%
%

\vspace{-2.5mm}
\section{Scalar Scheduling: Minimizing $q$-norms}
	\label{sec:scalar-q}
\vspace{-1.5mm}
As discussed earlier, our algorithm has two parts: a fractional algorithm that 
assigns jobs fractionally to machines, and a rounding algorithm that converts 
the fractional solution to an integer solution. We present the fractional algorithm
here, and defer the rounding algorithm to Section \ref{app:scalar-q}. We will assume 
throughout that we are working on a smoothed instance, which is wlog by 
Lemma~\ref{lem:canonical-gen}.

\newcommand{\cG}{\mathcal{G}}

To define the fractional algorithm, we first define a fractional relaxation of the 
$q$-norm objective. Let us use $G$ to index machine groups; let $|G|$ be the number
of machines in group $G$, $p_{Gj}$ be the processing time of job $j$ on any 
machine of group $G$, and $x_{Gj}$ be the fraction of job $j$ assigned to group $G$.
Also, let $s_G$ denote the speed of machines in group $G$.
The (fractional) load of a machine group $G$ is the ratio of the
total time for processing the fractional jobs assigned to the group
and the number of machines in the group: 
$$\Lambda_G = \sum_{j=1}^n \frac{1}{|G|} \cdot x_{Gj} p_{Gj},~\text{~where~} p_{Gj} = \frac{p_j}{s_G}.$$
Then, the fractional objective is:
\begin{equation}
	\label{eqn:group-view-scalar}
	h(x) := \sum_G |G| \cdot \left ( \Lambda_G \right)^q  
    + \sum_G \sum_j    (p_{Gj})^q \cdot x_{Gj}.
\end{equation}
The first term in $h(x)$ is simply the $q^q$-norm defined on the fractional 
loads, and the second term ensures that large jobs do not create a 
large integrality gap. We call these $f(x) := \sum_G |G| \left ( \Lambda_G \right)^q$
the {\em load-dependent} objective, $g(x) := \sum_G \sum_j    (p_{Gj})^q \cdot x_{Gj}$ 
the {\em job-dependent} objective, and their sum $h(x)$ the {\em total} objective of solution $x$. 

The goal of the fractional algorithm is to obtain a fractional solution $x$ that is 
$c^q$-competitive, for some constant $c$, for the total objective $h(x)$.




\smallskip
\noindent
{\bf Algorithm.} We use a (slightly modified) gradient descent algorithm defined for the 
objective $h(x)$. 
To define the algorithm, we denote the two terms in the derivative 
$\frac{d h(x)}{d x_{Gj}}$ by:
\begin{alignat*}{4}
\alpha_{Gj} &:=~ & \frac{d f(x)}{d x_{Gj}} & =~ |G|\cdot q \cdot \left(\Lambda_G\right)^{q-1}\cdot \frac{1}{|G|} \cdot p_{Gj} 
		~=~ q \cdot \left(\Lambda_G\right)^{q-1} \cdot \frac{p_j}{s_G} \\
\beta_{Gj} &:=~ & \frac{d g(x)} {d x_{Gj}} & =~ (p_{Gj})^q 
		~=~ \left(\frac{p_j}{s_G}\right)^q
\end{alignat*}
The algorithm assigns an infinitesimal fraction of the current job $j$ 
to the machine group $G$ that has the minimum value of 
$\eta_{Gj} := \max (\alpha_{Gj}, \beta_{Gj})$.
In case of a tie, the following rule is used:
\begin{itemize}
	\item If there is a tied machine group with $\alpha_{Gj} < \beta_{Gj}$,
    then this machine group is used for the assignment. Note that there 
    can only be at most one machine group with this property, by {\bf Property 3}
    of smoothed instances.
    \item If $\alpha_{Gj} \geq \beta_{Gj}$ for all tied machine groups, then
    we divide the infinitesimal job among the tied groups in proportion to
    $|G|\cdot s_G^\gamma$, where $\gamma = q/(q-1)$. These proportions are
    chosen to preserve the condition that the values of $\alpha_{Gj}$ remain
    tied. This is formally stated in Claim~\ref{clm:tied}, which 
	can be verified by a simple calculation that we defer to the appendix
    for brevity.
   \begin{claim}
\label{clm:tied}
	If a job $j$ is assigned in proportion to $|G|\cdot s_G^\gamma$ among machine groups
    $G$ with identical values of $\alpha_{Gj}$, where $\gamma = q/(q-1)$, then the value 
    of $\alpha_{Gj}$ remains equal for these machine groups after the assignment.
\end{claim}
\end{itemize}


\smallskip
\noindent
{\bf Analysis.} Our first lemma shows that at any point of time, the values of $\alpha_{Gj}$
for any job $j$ varies monotonically with the speed of the machine groups. 

\begin{lemma}
\label{lma:monotone}
	At any point of time, if $s_G > s_{G'}$, then $\alpha_{Gj} \geq \alpha_{G'j}$
    for any job $j$.
\end{lemma}
\begin{proof}
	First, note that the lemma holds for all jobs if it does for any single job.
    We now prove the lemma by showing that it inductively holds for the current 
    job $j$ at any time. For the property to be violated by the current fractional
    assignment, this assignment must be on group $G'$ with $\alpha_{Gj} = \alpha_{G'j}$.
    Now, note that $\beta_{G'j} > \beta_{Gj}$ by {\bf Property 3} of smoothed instances. 
    Therefore, the algorithm can make an assignment on $G'$ only if $G$ and $G'$
    are tied with 
    $$\eta_{Gj} = \alpha_{Gj} = \alpha_{G'j} = \eta_{G'j}.$$
    In this case, the algorithm assigns job $j$ to groups $G$ and $G'$ in proportion to
    $|G|\cdot s_G^\gamma$ and $|G'|\cdot s_{G'}^\gamma$, 
    where $\gamma = q/(q-1)$. This assignment preserves 
    $\alpha_{Gj} = \alpha_{G'j}$ by Claim~\ref{clm:tied}, 
    hence the lemma continues to hold.
\end{proof}

We fix an optimal solution $\opt$, and denote the fractional algorithm's solution 
by $\algo$; let the corresponding fractional assignments be $x_{\opt}$
and $x_{\algo}$. Let $\opt(j)$ (resp., $\algo(j)$) be the machine group on which a job 
$j$ is assigned by $\opt$ (resp., $\algo$). We call the assignment of a fractional job 
a {\em red assignment} if $\opt$ assigns $j$ on a slower machine group, i.e.,
if $s_{\opt(j)} < s_{\algo(j)}$; we call it a {\em blue assignment} if $\opt$
assigns $j$ on a faster machine group, i.e., $s_{\opt(j)} > s_{\algo(j)}$. If 
$\opt(j) = \algo(j) = G$, we call it a red assignment if $\beta_{Gj} \geq \alpha_{Gj}$ when
   the assignment was made; else, we call it a blue assignment. 


We will analyze 
the total increase in the objective $h(x_{\algo})$ caused by red and blue assignments 
separately. Note that there was a special case in the algorithm when machine 
groups were tied, where we assigned a fractional job to multiple machine groups.
However, in this case, by {\bf Property 2} of smoothed instances, at least half the 
job is assigned to the slowest tied machine group. Since $\eta_{Gj} = \alpha_{Gj}$
for all tied groups in this case, the increase in $h(x)$ overall is at most a constant
factor times the increase of $h(x)$ on the slowest machine group. Therefore, in this 
analysis, we will only consider the slowest machine group in this scenario. 

We first bound the contribution from red assignments.
\begin{lemma}
\label{lma:red}
	The total increase in $h(x_{\algo})$ due to red assignments of $\algo$
    is at most twice the job-dependent objective $g(x_{\opt})$ of $\opt$.
\end{lemma}
\begin{proof}
	Consider a red assignment of job $j$. We have two cases.
	First, suppose $s_{\opt(j)} < s_{\algo(j)}$.
    Given that we
    only consider the assignment on the slowest group in case of 
    a tie, we can conclude that:
    \begin{equation*}
     	\eta_{\opt(j)j} 
        > \eta_{\algo(j)j}
    	= \max(\alpha_{\algo(j)j}, \beta_{\algo(j)j}) 
    	\geq \alpha_{\algo(j)j} 
    	\geq \alpha_{\opt(j)j} ~\text{(by Lemma~\ref{lma:monotone})}.
    \end{equation*}
    Therefore,
    	$\beta_{\opt(j)j} > \alpha_{\algo(j)j}$.
    But, since $\beta_{\opt(j)j} > \beta_{\algo(j)j}$ as well, it 
    follows that
   	\begin{equation*}	
		\alpha_{\algo(j)j} + \beta_{\algo(j)j} 
		< 2 \beta_{\opt(j)j}.
	\end{equation*}
	
	Next, suppose $\opt(j) = \algo(j)$.
    In this case, 
	\begin{equation*}	
		\alpha_{\algo(j)j} + \beta_{\algo(j)j} 
		\leq 2 \max(\alpha_{\algo(j)j}, \beta_{\algo(j)j})
        = 2 \beta_{\algo(j)j} 
		= 2 \beta_{\opt(j)j},
	\end{equation*}
	where the second to last equality follows from the definition of red assignments. 
	To complete the proof of the lemma, we note that the increases in $g(x_{\opt})$
    are additive across all jobs.
\end{proof}

We are left to bound the total increase in $h(x_{\algo})$ due to blue assignments.
For blue assignments, $\opt$ assigns the fractional jobs to faster machine groups.
To understand the intuition behind our analysis of blue assignments, let us imagine
an idealized scenario where $\algo$ equalized the values of
$\alpha_{Gj}$ across all machine groups $G$ for all jobs $j$. In this case, 
$\algo$ produced an optimal assignment for the load-dependent objective.
Therefore, $f(x_{\algo}) \leq f(x_{\opt})$. The same argument works even if 
$\alpha_{Gj}$ is not equal for all groups, provided all jobs are blue,
by replacing uniformity of $\alpha_{Gj}$ by the monotonicity property from 
Lemma~\ref{lma:monotone}. However, there are two main difficulties with 
generalizing this argument further.
First, for a blue assignment of job $j$ to machine group $\algo(j)$, it may be 
the case that $\beta_{\algo(j)j} > \alpha_{\algo(j)j}$. In this case, bounding
the the load-dependent objective of $\algo$ is not sufficient. Second, 
we need to account for the fact that not all assignments are blue, and the monotonicity
guaranteed by Lemma~\ref{lma:monotone} might be contingent on red assignments.

To address the first issue, we specifically consider the blue assignments
with $\beta_{\algo(j)j} > \alpha_{\algo(j)j}$; let us call them {\em special} 
assignments. For all such special assignments, we modify
$\algo$ to $\algo'$ by additionally assigning the fractional job to 
the machine group (denoted $\algo(j)^+$) that is immediately 
faster than $\algo(j)$. The idea behind this addition is that 
$\alpha_{\algo(j)^+j} \geq \eta_{\algo(j)j}$ irrespective of
which of $\beta_{\algo(j)j}$ or $\alpha_{\algo(j)j}$
defines $\eta_{\algo(j)j}$.
Therefore, we can bound the increase in total objective due to special 
assignments by the increase in the load-dependent objective due
to the dummy assignments that we added.
Correspondingly, we modify $\opt$ to $\opt'$ by adding
a second copy of each such fractional job to $\opt(j)$.
Note that for special blue assignments, we have the strict inequality 
$s_{\opt(j)} > s_{\algo(j)}$; else, we would call it 
a red assignment. Hence, these additional dummy assignments are
also blue assignments.

We now show that these modifications do not significantly 
change the objectives of the respective solutions, while 
allowing us to only focus on the load-dependent objectives
$f(x_{\opt'})$ and $f(x_{\algo'})$. 
The first lemma is immediate.
\begin{lemma}
\label{lma:modify-4}
	The load-dependent objective $f(x_{\opt'})$ in $\opt'$ is at most 
    $2^q$ times the corresponding objective $f(x_{\opt})$ in $\opt$.
\end{lemma}

\begin{lemma}
\label{lma:modify-1}
	The total objective $h(x_{\algo})$ due to blue assignments in $\algo$ 
    is at most twice the load-dependent objective $f(x_{\algo'})$ due to 
    blue assignments in $\algo'$ .
\end{lemma}
\begin{proof}
	We consider two cases. 
	First, suppose $\alpha_{\algo(j)j} \geq \beta_{\algo(j)j}$. This is not
    a special blue assignment. In this case, 
    \begin{equation*}
    	\alpha_{\algo(j)j} + \beta_{\algo(j)j} \leq 2\alpha_{\algo(j)j}.
	\end{equation*}        
    Since $\algo'$ has at least as much load on every machine group as $\algo$,
	it follows that the total increase of objective in $\algo$ due to assignments 
	in this case is at most twice the load-dependent objective of $\algo'$. 
	
	Next, suppose $\alpha_{\algo(j)j} < \beta_{\algo(j)j}$ in a blue assignment.
    This is a special blue assignment, and we have $s_{\opt(j)} > s_{\algo(j)}$,
    as noted earlier. In this case, $\beta_{\algo(j)^+j} < \beta_{\algo(j)j}$, but
	$\eta_{\algo(j)^+j} \geq \eta_{\algo(j)j}$. Therefore, 
	$\alpha_{\algo(j)^+j} \geq \beta_{\algo(j)j}$ and 
	$\alpha_{\algo(j)^+j} \geq \alpha_{\algo(j)j}$.  Therefore, we have 
    \begin{equation*} 
    \alpha_{\algo(j)j} + \beta_{\algo(j)j} \leq 2\alpha_{\algo(j)^+j}.
    \end{equation*} 
    But, for every special assignment to 
	machine group $\algo(j)$ in $\algo$, there is a corresponding 
    assignment to machine $\algo(j)^+$ in $\algo'$.
	Therefore, the total increase of objective in $\algo$ due to 
    special assignments is at most twice the load-dependent objective 
    of $\algo'$. 
\end{proof}

Next, to handle our second issue, we  modify $\opt'$ to $\opt''$
by adding the load due to red assignments in $\algo$ on each machine.
This allows us to view the red assignments as blue assignments 
for the purposes of this analysis, since $\opt''$ now has a copy of 
every red job on the same machine as $\algo$.
Again, we establish that this transformation does not significantly
change the load-dependent objective of $\opt'$.

\begin{lemma}
\label{lma:modify-3}
	The load-dependent objective $f(x_{\opt''})$ in $\opt''$ is at most $2^q$ times the 
    load-dependent objective $f(x_{\opt'})$ in $\opt'$ plus $2^{q+1}$ times the job-dependent
    objective $g(x_{\opt})$ in $\opt$.
\end{lemma}
\begin{proof}
	We classify machine groups into two groups. The first type of group is one
    where the load in $\opt'$ is at least its load from red assignments in $\algo$. 
    The load in $\opt''$ for such groups
	is at most twice the load in $\opt'$. Therefore for these machine groups,
    the load-dependent 
	objective in $\opt''$ is at most $2^q$ times load-dependent objective 
	in $\opt'$. 
	
	The second type of machine group is one where the red load in $\algo$ is 
    more than the load in $\opt'$.  The load in $\opt''$ for such machine 
    groups is at most twice the red load in $\algo$. 
    Therefore by Lemma~\ref{lma:red}, the load-dependent objective in $\opt''$ 
    is at most $2\cdot 2^q$ times the job-dependent objective $g(x_{\opt})$ in $\opt$.
\end{proof}

We will now be able to apply our high level approach and show that the load-dependent objective of $\algo'$ is bounded by that of $\opt''$. We first show 
the following theorem on load profiles, which formalizes our earlier 
intuition. 

\begin{lemma}
\label{lma:load-profile}
	Consider two load profiles $\psi$ and $\xi$ over the machine groups with the 
	following properties:
	\begin{enumerate}
		\item (First condition) For any prefix $\cal G$ of machine groups in 
        decreasing order of speeds, the total job volumes
		satisfy: $\sum_{G\in {\cal G}} \psi_G \cdot |G| \cdot s_G \geq \sum_{G\in {\cal G}} \xi_G \cdot |G| \cdot s_G$.
		\item (Second condition) There exists a $\mu \leq 1$ such that
        for any two machine groups $G$ and $G'$, 
		we have: 
		$$\frac{\xi_G^{q-1}}{s_G} \geq \mu \cdot \frac{\xi_{G'}^{q-1}}{s_{G'}}.$$
	\end{enumerate}
	Then, the load-dependent objective of load profile $\psi$ is at least
	$\mu^{\frac{q}{q-1}}$ times 
	the load-dependent objective of load profile $\xi$. 
\end{lemma}
\begin{proof} 
	First, we transform the load profile $\xi$ to $\chi$ so as to 
    change the value of $\mu$ to $1$ in the second condition. For any
    group $G$, We set
    $\chi_G$ so that it satisfies
    $$\frac{\chi_G^{q-1}}{s_G} = \min_{G': s_{G'}\geq s_G} \frac{\xi_{G'}^{q-1}}{s_{G'}}.$$
    Since $\chi_G \leq \xi_G$ for any machine group $G$, the first 
    condition holds for $\psi$ and $\chi$ as well. Furthermore, by definition
    of $\chi$, it satisfies the second condition with $\mu = 1$. Finally,
    note that by the second condition on $\xi$,  
    \begin{equation}
    \label{eq:scale}
    	\frac{\chi_G^{q-1}}{s_G} 
    	= \min_{G': s_{G'}\geq s_G} \frac{\xi_{G'}^{q-1}}{s_{G'}} 
    	\geq \mu \cdot \frac{\xi_G^{q-1}}{s_G}.
    \end{equation}
    
	Now, we use an exchange argument to transform $\psi$ without increasing
	its  load-dependent objective until for every machine group $G$,
	we have $\psi_G \geq \chi_G$.
	In each step of the exchange, we identify the slowest machine group $G$
	where $\psi_G < \chi_G$. By the first condition, there must be a 
	machine group $G'$ with $s_{G'} > s_G$ such that $\psi_{G'} > \chi_{G'}$
    and for every prefix $\cal G$ of machine groups in decreasing order
    of speeds containing $G'$ but not containing $G$, the following 
    strict inequality holds:
    \begin{equation}
    \label{eq:first}
    \sum_{G\in {\cal G}} \psi_G \cdot |G| \cdot s_G 
    > \sum_{G\in {\cal G}} \chi_G \cdot |G| \cdot s_G.
    \end{equation}
    Furthermore, using the second condition (with now $\mu =1$), we have that 
	\begin{equation}
	\label{eq:prefix}
	\frac{\psi_{G'}^{q-1}}{s_{G'}} 
	> \frac{\chi_{G'}^{q-1}}{s_{G'}}
	\geq \frac{\chi_G^{q-1}}{s_G}
	> \frac{\psi_G^{q-1}}{s_G}.
	\end{equation}
	Now, we move an infinitesimal job volume from group $G'$ to group $G$ in $\psi$.
	Inequality \eqref{eq:prefix} implies that the load-dependent objective
	of $\psi$ decreases due to this move. Furthermore, both conditions of the 
	lemma continue to remain valid by Eqs.~\eqref{eq:first} and \eqref{eq:prefix}. 
    Such moves are repeatedly performed to obtain a load profile $\psi'_G$
    with at most the load-dependent objective of $\psi$, but additionally
    satisfying $\psi'_G \geq \chi_G$ for all machine groups $G$. 
    
    At this point,
	the lemma holds for the transformed load profile $\chi$ with $\mu = 1$. 
	To translate this back to the original load profile $\xi$, note that
    Eq.~\eqref{eq:scale} implies that $\chi_G \geq \mu^{1/(q-1)}\cdot \xi_G$
    for every machine group $G$. 
\end{proof}

We now apply Lemma \ref{lma:load-profile} to $\algo'$ and $\opt''$ 
to get our desired bound. 
\begin{lemma}
\label{lma:modify-2}
	The load-dependent objective of $\algo'$ is at most $2^q$ times 
	the load-dependent objective of $\opt''$.
\end{lemma}
\begin{proof}
	In Lemma~\ref{lma:load-profile}, we set $\psi$ to the load profile of 
    $\opt''$ and $\xi$ to the load profile of $\algo'$. 
	
	The first condition of Lemma~\ref{lma:load-profile} follows
	from the following observations:
	(a) for blue assignments in $\algo$, $s_{\opt(j)} \geq s_{\algo(j)}$;
	(b) for red assignments in $\algo$, the same fractional job $j$ is
    assigned to $\algo(j)$ in transforming $\opt'$ to $\opt''$;
    (c) finally, for special assignments added in transforming $\algo$
        to $\algo'$, we have $s_{\opt(j)} > s_{\algo(j)}$, i.e., 
        $s_{\opt(j)} \geq s_{\algo(j)^+}$. 
        
	We now check the second condition of Lemma~\ref{lma:load-profile}. From 
	Lemma~\ref{lma:monotone}, the condition holds with $\mu = 1$ for $\algo$.
	In $\algo'$, the load $\Lambda_{G^+}$ on a machine group $G$ increases by the 
    total load due to special assignments on machine group $G$, i.e., by at most
	$\Lambda_G\cdot \frac{s_G}{s_{G^+}} \leq \Lambda_G$. But, by Lemma~\ref{lma:monotone},
	$\Lambda_G \leq \Lambda_{G^+}$. Therefore, the load on machine group $G^+$ increases 
    by at most a factor of $2$. It follows that the second condition of 
    Lemma~\ref{lma:load-profile} holds with $\mu = 1/2^{q-1}$.
	
	Now, the lemma follows by applying Lemma~\ref{lma:load-profile}.
\end{proof}

Combining Lemmas \ref{lma:modify-4}, \ref{lma:modify-1}, \ref{lma:modify-3}, and 
\ref{lma:modify-2}, we obtain the desired bound for blue assignments:

\begin{lemma}
\label{lma:blue}
	The total increase in objective due to blue assignments in $\algo$ is at most 
	$a^q$ times the load-dependent objective of $\opt$, for some constant $a$.
\end{lemma}

Lemmas~\ref{lma:blue} and \ref{lma:red} imply that the algorithm is $c^q$-competitive 
on objective $h(x)$ for some constant $c$, as desired.

\vspace{-2.5mm}

\vspace{-2.5mm}
\section{Scalar Scheduling: Minimizing $q$-norms (Rounding)}
\label{app:scalar-q}
 \vspace{-1.5mm}

We presented the fractional algorithm for scalar scheduling for $q$-norms
in Section~\ref{sec:scalar-q}. In this section we give a rounding procedure that converts a fractional assignment to an integral assignment with a loss of $c^q$ for some constant $c$. This result in conjunction with the fractional algorithm from Section \ref{sec:scalar-q} implies a $(c\cdot b)^q$-competitive 
algorithm 
for optimizing the following objective.
\begin{equation}
h(x) :=  \sum_i \left(\sum_j x_{ij} p_{ij}\right)^q + \sum_{i,j} (p_{ij})^q x_{ij}.
\end{equation}
\medskip
\noindent
{\bf Rounding Algorithm.} Recall we can assume that machines have been smoothed wlog. 
It is straightforward to see that we can assume wlog that all machines in each group have identical fractional assignments of jobs. 
Since all machines in the same group are identical, we can focus on assignments at the granule of groups. In this spirit, we denote the fractional assignment of jobs to groups by $x_{G_lj} := \sum_{i \in G_l} x_{ij}$. Let $m(j)$, which we call $j$'s middle point, be the slowest group $G_l$ (as before, a group's speed is defined as that of any machine in the group) such that $j$ is processed by more than half on machines in groups $G_0, G_1, ..., G_l$, i.e. $\sum_{l \leq m(j)} x_{G_lj} \geq 1/2$; note that $\sum_{l \geq m(j)} x_{G_lj} \geq 1/2$.Then, we `commit' job $j$ to group $G_l$. Jobs committed to group $G_l$ are then scheduled greedily within the group 
(assigned to the machine with the smallest load).


\medskip
\noindent
{\bf Analysis.}
We show that committing job $j$ to its middle point group $G_{m(j)}$ and then using greedy algorithm to schedule the job within group $G_{m(j)}$, we only lose $O(1)^q$ factor w.r.t the objective. 

Consider any fractional solution $x^o$. Let $G_{m(j)}$ be the middle point group of job $j$ in $x^o$. Let's say that a solution/assignment is restricted if each job $j$ must be assigned to groups $G_0$, $G_1$, \ldots, $G_{m(j)}$. At a high-level, we  first show that this restriction can increase the objective by $O(1)^q$ factor. We then show that the further restriction that job $j$ can only go to machines in $G_{m(j)}$ can increase the objective by $O(1)^q$ factor. Let $x'$ denote a fractional assignment that is obtained from $x^o$ by doubling each job $j$'s assignment to groups $G_0, G_1, ..., G_{m(j)}$ (and discarding some assignments so that $\sum_{i} x'_{ij} = 1$), and $x''$ be a fractional assignment where each job $j$ is equally assigned to machines in $G_{m(j)}$.  

\begin{lemma}
	$h(x'') \leq O(1)^q h(x^o)$
\end{lemma}

For a formal proof, we decompose the objective.
\begin{align*}
h_1(x) &:= \sum_i \left(\sum_j x_{ij} p_{ij}\right)^q & h_2(x) &:= \sum_{i,j} p_{ij}^q x_{ij}  \\
\end{align*}

\begin{lemma}
\label{lem:First-part}
	$h_1(x'') \leq 2^q h_1(x') \leq 4^q h_1(x^o)$. 
\end{lemma}
\begin{proof}
Let $J_m$ denote the set of jobs with the same middle point $m$. If we only need to schedule jobs $J_m$, 	
	due to the optimality condition (see Claim~\ref{clm:tied}), 
we can see that  $\sum_i (\sum_{j \in J_m} p_{ij} x_{ij} )^q$ is minimized when for each $j \in J_m$, $x_{ij}$ is in proportional to $s_i^\gamma$ for all  machines $i$ in groups $G_0, G_1, ..., G_{m}$. Thus, when $x_{G_0 j} / S(G_0) = x_{G_1 j} / S(G_1) = ... = x_{G_m j} / S(G_m)$, where 
 $x_{G_t j} := \sum_{i \in G_t} x_{ij}$, as before. Knowing that $S(G_m) \geq S(G_0) + S(G_1) + S(G_2) + ... + S(G_{m-1})$ by (at most) doubling the assignments to  $G_m$, we can fully assign jobs in $J_m$ to (machines in) $G_m$. This will only increase the objective by a factor of $2^q$. Further, no two jobs with different middle points are assigned to the same group. This proves the first inequality. 
The second inequality follows since each machine's load at most doubles when we convert $x^o$ into $x'$.
%
\end{proof}

\begin{lemma}
\label{lem:second-part}
	$h_2(x'') \leq 2 h_2(x^o)$. 
\end{lemma}
\begin{proof}
	Fix a job $j$. Any machine $i \in G_{j(m)}$ is faster than any machine $i'$ in $G_{j(m)} \cup G_{j(m)+1} \cup \cdots$. Thus, $p_{ij} \leq p_{i'j}$, hence we can charge $j$'s contribution to the second term in $x''$ to $j$'s contribution to the second term in $x^o$ on machines  in 
$G_{j(m)} \cup G_{j(m)+1} \cup \cdots$. The factor 2 follows since $j$ is assigned to machines in $G_{j(m)} \cup G_{j(m)+1} \cup \cdots$ by at least half. 
\end{proof}
To complete the analysis, it suffices to show that the integral solution $\overline{x}$ produced 
by the greedy algorithm is $c^q$-competitive against $h(x'')$ for some constant $c$.

\begin{lemma}
$h(\overline{x}) \leq (2^q +1)h(x'')$
\end{lemma}
\begin{proof}
Fix a group $G_l$, and let $h_l(x)$ be the objective for just group $l$.
Let $\widehat{p_i}$ be the load of the last job that was assigned to machine $i$,
and let $\Lambda_i'$ be the load on machine without this last job (i.e., $\Lambda_i' = \Lambda_i - \widehat{p_i}$). Let $\algo(j)$ be the machine to which $j$ is assigned by the greedy algorithm. 
 Observe that  
\begin{alignat*}{2} 
h_l(\overline{x}) &= \sum_{i \in G_l}\left(\sum_j p_{ij}\overline{x_{ij}} \right)^q + \sum_{i \in G_l}\sum_j \overline{x_{ij}}p_{ij}^q \\	
 & = \sum_{i \in G_l}\left(\Lambda'_i + \widehat{p_i} \right)^q + \sum_j p_{\algo(j)j}^q \\
 & \leq \sum_{i \in G_l}\left(2 \max(\Lambda'_i, \widehat{p_i}) \right)^q + \sum_j p_{\algo(j)j}^q \\
 & \leq 2^q\sum_{i \in G_l}\left((\Lambda_i')^q + \widehat{p_i}^q\right) + \sum_j p_{\algo(j)j}^q \\
 & \leq (2^q + 1)\left(\sum_{i \in G_l}(\Lambda_i')^q + \sum_j p_{\algo(j)j}^q \right) \leq (2^q + 1)h_l(x'').\\
\end{alignat*}
The last inequality follows since $x''$ assigns all jobs within a group evenly (i.e. $x_{ij}'' = 1/|G_l|$
for all $i$ in the group); therefore, since the algorithm assigns greedily, $\sum_{i \in G_l}(\Lambda_i')^q$
is bounded by $\sum_{i \in G_l}\left(\sum_{j}x_{ij}''p_{ij}\right)^q$. Similarly, $\sum_j p_{\algo(j)j}^q$ is is equal to $  \sum_{i \in G_l}\sum_j x_{ij}''p_{ij}^q$ since all machines have identical speeds within the group.

Summing the bound over all groups $l$, we obtain that $h(x'') \leq (2^q + 1)h(\overline{x})$. 
\end{proof}

\section{Vector Scheduling: Minimizing $q$-norms}
	\label{sec:vector-q}
\vspace{-1.5mm}
\newcommand{\uniform}{\textsc{uniform}}
\newcommand{\bluejob}{\textsc{blue}}
\newcommand{\redjob}{\textsc{red}}
\newcommand{\greyjob}{\textsc{grey}}
\newcommand{\optr}{\mathsf{opt_r}}
\newcommand{\opts}{\mathsf{opt-s}}
\newcommand{\optG}{\mathsf{opt_G}}

As in the previous section on scalar scheduling, we present our fractional algorithm
for vector scheduling here, and defer the rounding algorithm to Section~\ref{app:vector-q}. 
In this section we will obtain a fractional solution that is $O(\log^2d)$-competitive. 
Then, using the rounding algorithm in Section~\ref{app:vector-q}, we will round it with a loss of $O(\log d / \log \log d)$ factor in the competitive ratio, thus proving the first part of Theorem~\ref{thm:related-pnorm}.  We  assume that $q \geq \log d$ since otherwise we can use the any-norm-minimization algorithm for unrelated machines in \cite{ImKKP15} to find a $O(\log d + q)$-competitive solution. We further assume that $q > 1$ since if $q =1$, assigning all jobs to the fastest machines yields an optimal solution. 

   \subsection{Overview of Algorithm and Analysis}

In this section, our goal will be to find a fractional solution that is $O(\log^2d)^q$ competitive against the following objective:
\begin{equation}
	\label{eqn:main-obj-nogroup}
	 \sum_i \sum_k \Big(\sum_j p_{ij}(k) x_{ij} \Big)^q + \sum_{i,j,k} \Big(p_{ij}(k) x_{ij} \Big)^q, 
\end{equation}
where $p_{ij}(k)$ denotes $p_j(k)/ s_i$. We first argue that this objective is valid, i.e., if the algorithm is competitive 
on this relaxation then the algorithm is competitive for our original objective of minimizing the maximum $q$-norm across 
all dimensions. 

\begin{lemma}
	\label{lem:obj-justification}
    An algorithm
that is $O(\gamma)^q$-competitive with respect to  
objective \eqref{eqn:main-obj-nogroup} (which sums over all dimensions)
implies the algorithm is $O(\gamma)$-competitive for our
desired objective stated in the introduction (optimizing for the maximum $q$-norm across 
all dimensions; call this the original objective). 
\end{lemma}
\begin{proof}
Recall our definitions of load-dependent, job-dependent, and total objective from Section \ref{sec:scalar-q}. 
Let $\|\Lambda^*(k)\|_q$ denote the $q$-norm of the $k$th dimension in the optimal solution.   Clearly
the optimal total objective in a fixed dimension $k$ is within a $O(1)^q$ factor of $\|\Lambda^*(k)\|_q^q$ (since 
the job-dependent objective is a lower bound on $\|\Lambda^*(k)\|_q^q$). We also have that the 
optimal solution to objective \eqref{eqn:main-obj-nogroup} is at most $d$ times $\|\Lambda^*(k')\|_q^q$, 
where $k'$ is the dimension with the maximum $q$-norm. However, since we assume that $q \geq \log d$, we have 
that $d \leq 2^q$. Thus, putting these observations together, we have that optimal solution to \eqref{eqn:main-obj-nogroup}
is at most $O(1)^q$ times the optimal solution to the original objective, implying the a $O(\gamma)^q$ competitive algorithm 
for this relaxation is $O(\gamma)$-competitive on the original objective. 
\end{proof}

As before, we also preprocess machines to create a smoothed instance, which is wlog by Lemma~\ref{lem:canonical-gen}. Thus our the objective we will use is the following: 

\begin{equation}
	\label{eqn:main-obj}
	\sum_k \sum_G |G| \left ( \frac{1}{|G|} \sum_j p_{Gj}(k) x_{Gj} \right)^q  + \sum_G \sum_j     \left(\sum_k  (p_{Gj}(k))^q \right) x_{Gj},
\end{equation}
where $x_{Gj}$ denotes that fraction of job $j$ assigned to group $G$. Recall that within a given group $G$, 
we can assume that all jobs assigned to $G$ are spread evenly among the machines in $G$. 

To simplify our presentation, we will assume that each job only has an infinitesimal fraction that needs assigned; namely, we will assume that job $j$ is fully assigned when $\sum_{i}  x_{ij} = \delta$ for an infinitesimally small value $\delta >0$. This modification can be done by replacing each job $j$ by a set of jobs $j_1, j_2, ..., j_{1 / \delta}$ with vector entries $\delta p_j(k)$ for each dimension $k$ and requiring that $\sum_{i} x_{ij_r} = \delta$ for these newly created jobs. Note that this alternate view does not change the objective considered by the algorithm or how the algorithm works since the algorithm is already making a fractional assignment.

We are now ready to present our algorithm. At a high level, the algorithm assigns each job in two phases.
In the first phase, we define a single scalar load derived from the job's maximum load entry and assign it using the 
scalar algorithm for $q$ norms given in Section \ref{sec:scalar-q}. This produces a fractional assignment which we will call the {\em scalar solution}. Using the scalar solution, we then determine a set of {\em candidate groups} $\cG_j$ to which job $j$ can go to in the second phase, i.e.,we only consider assignments where each job $j$ can only go to a group in $\cG_j$; call such assignments {\em restricted assignments}. A key Lemma, which we prove in Section  \ref{sec:step1}, is the following:

\begin{lemma}
	\label{lem:candidate-group}
    	The optimal fractional restricted assignment is at most $O(1)^q$ times the optimal assignment 
        
        with respect to objective \eqref{eqn:main-obj}. 
\end{lemma}

Thus, in the second phase, we produce an fractional (vector) assignment that is $O(\log^2 d)^q$-competitive against the optimal restricted assignment, which by Lemma \ref{lem:candidate-group} gives us an assignment with the desired competitive ratio. We now describe these two phases in more detail.


\paragraph{Phase 1: Producing the scalar assignment.} 
Let $p_{j, \max} := \max_k p_j(k)$. 
To define our scalar instance, we set scalar size of job $j$ to be $p_{j, \max} / d^2$.
Thus to schedule jobs in this phase, we simply use the algorithm for scalar loads from Section \ref{sec:scalar-q}.

Let $G_{f(j)}$ be the slowest group where $j$ is assigned in the scalar solution, and let $M$ be the number of groups.
Define:
\begin{equation*}
\cG_j :=\{G_{ \max\{0, f(j) - 4 \log d\} }, G_{\max\{0, f(j) - 4 \log d\} +1}, ... , G_{\min\{M, f(j) + 4 \log d\}}\},
\end{equation*}
which we call the candidate groups of job $j$.
In other words, $\cG_j$ is a collection of $O(\log d)$ consecutive groups containing $G_{f(j)}$ along with (potentially) some slower and some faster groups. Later in Lemma~\ref{lem:candidate-group}, we will show that there is a $O(1)^q$-approximate assignment w.r.t. (\ref{eqn:main-obj}) where each job $j$ is only assigned to groups in $\cG_j$.

\paragraph{Phase 2: Producing the restricted assignment.} In this phase, we produce a restricted assignment
assignment that is $O(\log^2 d)^q$-competitive against the optimal restricted assignment $\optr$,
which by Lemma \ref{lem:candidate-group} implies a $O(\log^2 d)^q$-competitive solution 
against the actual optimal solution. 
To do this, we maintain $O(\log d)$ separate sub-instances, each one 
corresponding to a set of disjoint candidate groups. Namely, let $\cG_G$ denote the set of
jobs $j$ such that $f(j) = G$ (i.e., the set of jobs whose candidate groups are centered 
around $G$).  There will $8 \log d + 1$ instances $0, \ldots, 8\log d$, where in the $t$th instance, 
we schedule jobs with candidate groups $\{\cG_t, \cG_{t + 8\log d + 1}, \cG_{t + 16\log d + 2}, \ldots\}$. 
It is not hard to verify that each set of candidate groups belongs to a unique instance, and the set of candidate groups 
within an instance are disjoint. 

Within each sub-instance, we will schedule jobs with the same candidate groups separately. Namely, 
fix a set of candidate groups $\cG$ and let $\mathsf{opt}_{\cG}$ 
be the optimal solution (and value of the optimal solution)
with respect to objective \eqref{eqn:main-obj} for scheduling just jobs 
with candidate groups $\cG$. Our goal will be to find a solution 
that satisfies the following set of constraints:

\begin{align}
\max_k \max_{G \in \cG}  \sum_j \frac{1}{|G|} p_{Gj}(k) x_{Gj} &\leq \mathsf{opt}_{\cG}^{1/q} \textnormal{ and }  \label{eqn:makespan-optimization}\\
\max_{G \in \cG}  \sum_j     \left(\sum_k  (p_{Gj}(k))^q \right) x_{Gj} &\leq \mathsf{opt}_{\cG}  \nonumber 
\end{align}
Note that $\mathsf{opt}_\cG$ satisfies these conditions. Also note that we will assume that $\mathsf{opt}_{\cG}$ is 
known from the outset of the instance (this assumption can be removed by using a standard doubling technique where 
the algorithm maintains a guess for $\mathsf{opt}_{\cG}$ and updates the guess by a factor of $2^q$ every time 
it is wrong; however for simplicity, we will assume $\mathsf{opt}_{\cG}$ is known for each set of candidate groups $\cG$).

We interpret this online problem as the makespan minimization for unrelated machines, i.e., we think of each group $G$ as a meta machine and of each job $j$ as having an averaged load $\frac{\delta p_{Gj}(k)}{|G|}$ on a meta-machine $G$ on dimension $k$.
We also create a special dimension 0 to encode the second set of constraints, where job $j$ has load $ \sum_k (\delta p_{Gj}(k))^q$ on meta-machine $G$ on dimension 0. Then, the problem is now reduced to finding an assignment where the makespan on dimension 0 is upper bounded by $\mathsf{opt}_{\cG}$, and the makespan on other dimensions from 1 to $d$ is upper bounded by $\mathsf{opt}_{\cG}^{1/q}$.  In \cite{ImKKP15}, this problem was studied under the name of any norm minimization for unrelated machines (VSANY-U). Using the algorithm in \cite{ImKKP15}, one can find a solution minimizing the $\log (O(|\cG|)$-norm on each dimension with the target values $\opt_{\cG}^{1/q}$ on dimensions $1, 2, 3, ... d$, and $\opt_{\cG}$ on dimension 0, which is equivalent to the makespan optimization problem defined by \eqref{eqn:makespan-optimization} up to a constant factor. 

This completes the description of the algorithm for Phase 2. We now show that the Phase 2 assignment is $O(\log^2 d)^q$-competitive 
  ainst the optimal restricted assignment $\optr$. First we argue that the solution produced in each sub-instance
is $O(\log d)^q$-competitive against $\optr$. 

\begin{lemma}
\label{lem:subinstance}
Fix a sub-instance $S$ from Phase 2. The objective of the solution produced by the algorithm for $S$ is at most $O(\log d)^q$ times that of the optimal restricted assignment $\optr$. 
\end{lemma}
\begin{proof}

First, fix a set of candidate groups $\cG$ in $S$, and consider the solution produced by the
VSANY-U algorithm given in \cite{ImKKP15} for $S$. This algorithm is $O(\log d + \log m)$-competitive, 
where $m$ is the number of machines. In our setting, the number of meta machines is $m = |\cG| = O(\log d)$,
and thus this algorithm will produce a solution such that the constraints in \eqref{eqn:makespan-optimization} 
are violated up to a $O(\log d + \log\log d) = O(\log d)$ factor. Thus is follows that this solution (denote it $\mathsf{algo}_\cG$) with respect to objective \eqref{eqn:main-obj} is at most:

\begin{equation*}
\mathsf{algo}_\cG = |\cG|  \cdot d \cdot (O(\log d)  \cdot \opt_\cG^{1 / q})^q +  |\cG| \cdot O(\log d)\cdot \opt_\cG  = O(\log d)^q \cdot \opt_\cG,
\end{equation*}
since $q \geq \log d$. 

Next, observe that since the candidate groups within a sub-instance are disjoint, we have that the algorithm's overall 
objective in the sub-instance (denote this $\mathsf{algo}_S$) equals $\sum_{\cG \in S}\mathsf{algo}_\cG$. Also, 
again since candidate groups are disjoint, we have $\sum_{\cG \in S} \opt_\cG \leq \optr$. Thus is follows 
that 

\begin{equation*}
\mathsf{algo}_S = \sum_{\cG \in S}\mathsf{algo}_\cG = \sum_{\cG \in S} O(\log d)^q \cdot \opt_\cG \leq O(\log d)^q \opt_r.
\end{equation*}
\end{proof}

Finally, we argue that the overall solution $\mathsf{algo}$ (i.e., combining the solutions produced over 
all sub-instances) is at most $O(\log^2 d)^q \cdot \optr$.

\begin{lemma}
The solution produced by Phase 2 is at most $O(\log^2 d)^q$ times the optimal restricted assignment. 
\end{lemma}

\begin{proof}
Let $T = O(\log d)$ denote the number of sub-instances. The overall objective that sums over all sub-instances $S$ can be bounded as follows:  
\begin{alignat*}{2}
\mathsf{algo} & = \sum_k \sum_G |G| \left (  \sum_S\frac{1}{|G|}\sum_{j \in S} p_{Gj}(k) x_{Gj} \right)^q  + \sum_S \sum_{G, j \in S} \left(\sum_k  (p_{Gj}(k))^q \right) x_{Gj} \\
& \leq  \sum_k \sum_G |G| \left ( T\cdot \max_S\left(\frac{1}{|G|}\sum_{j \in S} p_{Gj}(k) x_{Gj} \right)\right)^q  + \sum_S \sum_{G, j \in S} \left(\sum_k  (p_{Gj}(k))^q \right) x_{Gj} \\
& \leq T^q \sum_S \sum_k \sum_G |G|\left(\frac{1}{|G|}\sum_{j \in S} p_{Gj}(k) x_{Gj} \right)^q  + \sum_S \sum_{G, j \in S} \left(\sum_k  (p_{Gj}(k))^q \right) x_{Gj}. \\ 
& \leq T^q \sum_S \mathsf{algo}_S \leq T^q \sum_S O(\log d)^q \optr = O(\log^2 d)^q \cdot \optr,
\end{alignat*}
as desired. Note that the the last inequality follows by Lemma \ref{lem:subinstance}, and the last equality follows since the are $O(\log d) = O(1)^q$ sub-instances. 

\end{proof}

\subsection{Proof of Lemma \ref{lem:candidate-group}}

\label{sec:step1}

This section is devoted to showing Lemma \ref{lem:candidate-group}. Recall that $p_{j, \max} := \max_k p_j(k)$. We first observe that we can assume w.l.o.g. that each job $j$ has size at least $\frac{1}{d^2} p_{j, \max}$ on all dimensions.  

\begin{lemma}
	If we increase each job $j$'s load so that $j$ has load on dimension $\max \{p_j(k), \frac{1}{d^2} p_{j, \max}\}$,objective (\ref{eqn:main-obj}) increases by a factor of at most $2^q$.
\end{lemma}
\begin{proof}
	Consider any aggregate load vector on a fixed machine $i$, $\langle L_1, L_2, ...., L_d \rangle$. Consider an arbitrary dimension, say dimension 1. After the change, $L_1$ can increase up to $L_1 + \frac{1}{d^2} (L_2 + L_3 + ... + L_d)$. Thus, $(L_1 + \frac{1}{d^2} (L_2 + L_3 + ... + L_d))^q \leq 2^q (L_1)^q + \frac{2^q}{d^q} ((L_2)^q + ... + (L_d)^q)$. So one dimension can increase other dimension $k$'s contribution to the objective by only $2^q/ d$ times $k$'s contribution before the change. Hence the lemma follows. 	 
\end{proof}
    Thus we can assume w.l.o.g.\ that we run our algorithm after making this change to each job upon arrival. We note that this change is not necessary for the analysis, but it will help simplify our presentation. 

\newcommand{\unimax}{\textsf{unimax}}
\newcommand{\unimin}{\textsf{unimin}}

\smallskip
Consider an optimal schedule $\opt$ and the optimal restricted assignment $\optr$.  
Again to simplify the notation, we let $\opt$ and $\optr$ also denote their objective values, depending on context. We say that a job $j$ is red if it is assigned to 
a group not in $\cG_j$ that is slower than groups in $\cG_j$; similarly, the job is said to be blue if it is assigned to a group not in $\cG_j$ that is faster than groups in $\cG_j$; otherwise, the job is grey. We decompose the objective to analyze the contribution of jobs of each type, separately. In particular, let $\bluejob$, $\redjob$, $\greyjob$ denote set of blue, red, and grey jobs, respectively. Also denote $\optr^\bluejob$, $\optr^\redjob$ , and $\optr^\greyjob$ denote the optimal restricted assignments (and values) that just schedule blue, red, and grey jobs, respectively.


Observe that since grey jobs are scheduled on the same set of machines in both $\opt$ and $\optr^\greyjob$, we have that $\optr^\greyjob \leq \opt$.
Thus, the following decomposition is immediate.     
  
\begin{lemma}
	\label{lem:11}
	$\optr \leq 3^q (\optr^\bluejob  + \optr^\redjob + \opt)$.
\end{lemma}

Henceforth,  we will focus on bounding $\optr$ for red and blue jobs. The key idea is to reduce the problem to a single dimensional case. But this reduction is not free -- $\optr$ will have to deal with red and blue jobs of factor $d$ larger sizes than $\opt$. We will still be able to show that $\opt$ is considerably large compared to $\optr$ since $\opt$ processes jobs in groups that are so `out of range.'  From now on, 
we only consider red or blue jobs. 

\smallskip
We say that an input is uniform if every job has an equal size over all dimensions. We will consider two uniform inputs derived from the original input. Let $J^\unimax$ denote the set of jobs where each job $j$'s size vector is replaced with $p_{j, \max} \cdot \langle 1, 1, ..., 1 \rangle$. Similarly, let $J^\unimin$ denote the set of jobs where each job $j$'s size vector is replaced with $\frac{p_{j, \max}}{d^2} \cdot \langle 1, 1, ..., 1 \rangle$. Note that $J^\unimax$ is as hard as the original input, and $J^\unimin$ is as easy as the original input. Since our goal is to upper bound $\optr$ by $\opt$, we can safely assume that $\optr$ has to process $J^\unimax$ while $\opt$ does $J^\unimin$. Since all jobs have uniform sizes, all dimensions have an equal contribution to the objective. Hence, we can focus on an arbitrary dimension, and ignore all other dimensions. Accordingly, we can now assume that jobs have scalar sizes. 

  To recap, there are only red or blue jobs. And each job $j$'s size is $p_j / d^2$ for $\opt$ but $p_j$ for $\optr$; to simplify the notation we use $p_j$ in place of $p_{j, \max}$. Note that for each job $j$, $\optr$ assigns it to groups in $\cG_j$, but $\opt$ does to other groups. To compare $\optr$ to $\opt$, we assume that $\optr$ assigns each job $j$ to group $G_{f(j)}$.  Since this is a further restriction to $\optr$, we can safely assume. Recall that $G_{f(j)}$ is the group where the single dimensional case algorithm assigns job $j$ with scalar size $p_{j} / d^2$. To make our analysis more transparent, for each job $j$ we only keep job $j$'s assignment to 
$G_{f(j)}$. This is justified since  $j$ is assigned to $G_{f(j)}$ by at least half (of its portion $\delta$) as we observed
in Section~\ref{sec:scalar-q}. To factor in this, we will lose factor $2^q$.

\smallskip
Our remaining goal is to upper bound $\optr^\bluejob$ and $\optr^\redjob$ by $\opt$. 
We let $J_f$ denote the set of jobs assigned to $G_f$ in $\optr$. 

\begin{lemma}
	\label{lem:12}
	$\optr^\redjob \leq \opt$.
\end{lemma}
\begin{proof}
	Fix a group $f$.  Consider any job $j \in J_f$. The job was assigned to $G_f$, but not to any slower groups since $\alpha_{G_f j} < \beta_{G_{f+1} j}$. Hence the contribution of red jobs to $\optr$'s total objective is at most 
 \begin{equation*}
\optr(\redjob, f) := \delta \sum_{j \in G_f \cap \redjob} \left(\frac{p_j}{s_{G_{f+1}}}\right)^q.
\end{equation*} 

  Knowing the fastest group $\opt$ can use to process $j$ is $G_{f + 4 \log d +1}$, and its speed is at most $1 / d^4$ times that of $G_{f+1}$, $\opt$'s job-dependent objective for jobs in $G_f \cap \redjob$ is at least \\$\delta \sum_{j \in G_f \cap \redjob} (\frac{p_j / d^2 }{s_{G_{f+4 \log d + 1}}})^q \geq \optr(\redjob, f)$. Summing over all $f$, we have the lemma. 	
\end{proof}

\begin{lemma}
	\label{lem:13}
  	$\optr^\bluejob \leq \opt$.
\end{lemma}
\begin{proof}
  		Fix a group $f$. Consider blue jobs assigned to $G_f$ in $\optr$. As we observed in Section~\ref{sec:scalar-q}, if $G_f$ is the slowest group to which the single dimensional case algorithm assigns $j$, then we know that $\alpha_{G_{f-1}j} \geq \max \{\alpha_{G_f j}, \beta_{G_f j}\}$. Hence we can upper bound $\optr$'s total objective for jobs $J_f \cap \bluejob$ by $\optr(\bluejob, f) := |G_{f-1}| (\frac{V}{|G_{f-1}| s_{G_{f-1}}})^q = \frac{V^q}{(S(G_{f-1}))^{q-1}}$ where $V := \sum_{j \in J_f \cap \bluejob} p_j$. 		
		We know that $\opt$ can only use groups $G_0$, $G_1$, $G_2$, \dots, $G_{f - 4 \log d -1}$ to process jobs in $G_f \cap \bluejob$.
Let $T := f- 4 \log d -1$. 
		Now we would like to lower bound $\opt$ by only considering its load-dependent objective. Thus, we would like to  minimize the load-dependent objective when we're asked to process jobs of total size  $V / d^2$ only using groups $G_0, G_1, G_2, ..., G_{T}$. In other words, we would like to minimize
		 $\sum_{1 \leq t \leq T} |G_t| (\frac{V_t}{|G_t| s_{G_t}})^q = \sum_{1 \leq t \leq T} (\frac{V_t^q}{(S(G_t))^{q-1}})$ 
subject to $\sum_{1 \leq t \leq T} V_t = V / d^2$.  By an easy algebra, we can see that the minimum is $\frac{V}{ d^2} \eta^{q-1}$ where $\eta := \frac{V / d^2}{\sum_{1 \leq t \leq T} S(G_t)} \geq d^2 \cdot \frac{V}{S(G_{m-1})}$. Thus, $\frac{V}{ d^2} \eta^{q-1} \geq \optr(\bluejob, m)$ due to Properties 2 and 3; recall that $q$ is an integer greater than 1. By summing over all $f$, we have the lemma. 
 \end{proof}
		 
Thus we have proven Lemma~\ref{lem:candidate-group}.

\vspace{-2.5mm}
\section{Vector Scheduling: Minimizing $q$-norms (Rounding)}
\label{app:vector-q}
\vspace{-1.5mm}


In this section we give a rounding procedure that converts a fractional assignment to an integral assignment with a loss of 
$O( \frac{\log d}{\log \log d})$ 
factor in the competitive ratio for minimizing the $q$ norm when machines have homogeneous speeds. 
We will use the following objective, which is equivalent to our original objective up to a constant factor; see Lemma~\ref{lem:obj-justification}.

\begin{equation}
\label{eqn:main-obj_}
	h(x) := \max_k \sum_i  \left(\sum_i p_{ij}(k) x_{ij} \right)^q  +   \sum_{i,j} (p_{ij}(k))^q x_{ij}   
\end{equation}

\paragraph{Rounding Algorithm.}
Like scalar scheduling, since all machines in the same group are identical we focus on assignment of jobs to groups. We define $x_{G_lj}$ and $m(j)$ as before. We `commit' job $j$ to the middle point group $G_{m(j)}$. Then, we schedule job $j$ Jobs on one  of machines of this group by following the 
$O( \frac{\log d}{\log \log d})$-competitive algorithm for vector identical machines.

\paragraph{Analysis.}
We first show that we can commit each job $j$ to its middle point group, $G_{m(j)}$ without losing more than $O(1)^q$ factor w.r.t .(\ref{eqn:main-obj_}). We define $x^o$, $x'$ and $x''$ the same as previous section.

\begin{lemma}
	\label{lem:commit-to-a-group}
	 $h(x'') \leq O(1)^q h(x^o)$.
\end{lemma} 

To prove this lemma let's decompose the objective. Note that $h(x)$ and $h_1(x) + h_2(x)$ are within factor 2. 
\begin{align*}
h_1(x) &:= \max_k h_{1, k}(x)       & \mbox{ where } h_{1, k}(x) &:= \sum_i (\sum_j x_{ij} p_{ij}(k))^q \\
h_2(x) &:= \max_k  h_{2,k}(x) 	& \mbox{ where } h_{2, k}(x) &:= 	\sum_{i,j}  (p_{ij}(k))^q x_{ij} 
\end{align*}

 \begin{lemma}
	For any $x$, $h(x) \leq h_1(x) + h_2(x)$.
\end{lemma}
\begin{proof}
	Immediate from the definition of $h$, $h_1$ and $h_2$. 
\end{proof}

\begin{lemma}
	$h_1(x'') \leq 2^q h_1(x') \leq 4^q h_1(x^o)$. 
\end{lemma}
\begin{proof}
	Fix a dimension $k$. Consider scalar scheduling in this dimension. From the lemma~\ref{lem:First-part}, we can say $h_{1,k}(x'') \leq 2^q h_{1,k}(x') \leq 4^q h_{1,k}(x^o)$ for each $k$. Definition of $h_1(x)$ follows the lemma.
\end{proof}

\begin{lemma}
	$h_2(x'') \leq 2 h_2(x^o)$. 
\end{lemma}
\begin{proof}
	For each $k$, with the same argument as lemma~\ref{lem:second-part}, we have $h_{2,k}(x'') \leq 2 h_{2,k}(x^o)$. The lemma follows from definition of $h_2$. 
\end{proof}

From the above lemmas, the desired Lemma~\ref{lem:commit-to-a-group} follows. 


\bigskip

It now remains to show that given a fractional assignment where each job is assigned to only one group consisting of identical machines,  we can convert it into an integral assignment online using a  $O( \frac{\log d}{\log \log d})$-competitive algorithm for $d$-dimensional identical machines. Our goal is to establish a competitive ratio of $O( (\frac{\log d}{\log \log d})^q)$ against $h(x)$ when all machines are identical. Let $m$ denote the number of machines. 

Although \cite{ImKKP15} gives a $O( (\frac{\log d}{\log \log d})^{\frac{q-1}{q}})$-competitive algorithm for the $q$ norm, here we only present an online rounding algorithm that loses a competitive ratio of  $O( \frac{\log d}{\log \log d})$. The reason we present a slightly worse competitive ratio is because we need to argue against the objective $h(x)$, hence we can't do some part of the preprocessing done in \cite{ImKKP15}. Also  since we already lose an additional $O(\log^2 d)$ factor in other places, we choose not to further optimize this ratio. 

The rounding algorithm we use here is essentially the $O(\log d / \log \log d)$-competitive makespan minimization algorithm for identical machines \cite{ImKKP15}. Let's use the objective (\ref{eqn:main-obj}). As discussed before, this is equivalent to $h(x)$ up to a constant factor for minimizing the $q$ norm.  By a standard doubling trick, we can assume w.l.o.g. that we know the final maximum average load on any dimension, i.e., $A := \max_k \sum_{j } p_j(k) / m$. Note that the first term in the objective is lower bounded by $m A^q$; since all machines are identical, we assume w.l.o.g. that the speed is 1.  We say that a job $j$ is big on dimensions $k$ if $p_{ij}(k) \geq A$. Our rounding algorithm ensures that every machine gets at most $\eta$ big jobs on any dimension, and its total load of small jobs is at most $\eta A$ where $\eta  = O(\log d / \log \log d)$ where an appropriate constant is hidden. This can be  done by a independent rounding, followed by a postprocessing that takes are of `overloaded' jobs. The idea is, using standard concentration inequalities, to show that only a very small fraction of jobs need to be `reassigned' in the postprocessing.  This randomized rounding can be derandomzied using a potential function argument.

To see that this guarantee is sufficient to establish the desired competitiveness w.r.t. the objective (\ref{eqn:main-obj}),  consider any fixed machine $i$. Let $J_s$ and $J_b$ denote small and big jobs assigned to $i$, respectively. Let $X_{ij}$ denote a binary variable such that $X_{ij} = 1$ if and only if $j$ is assigned to machine $i$ after the rounding. 
Then, machine $i$'s contribution to the first term in the objective is, 
\begin{align*}
   	& \sum_k (\sum_j p_{j}(k) X_{ij} )^q  \\
 \leq & 2^q \sum_k (\sum_{j \in J_s} p_{j}(k) X_{ij})^q + 2^q \sum_k (\sum_{j \in J_b} p_{j}(k) X_{ij})^q \\
 \leq & 2^q \sum_k (\eta A)^q + 2^q \sum_k \eta^q \sum_{j \in J_b} (p_{j}(k))^q X_{ij},
\end{align*}
where the last inequality follows since each machine contains at most $\eta$ big jobs and each machine has at most $\eta A$ load of small jobs on any dimension $k$. Summing over all machines, we have 
\begin{align*}
   	& \sum_i \sum_k (\sum_j p_{j}(k) X_{ij} )^q  \\
 \leq & (2\eta)^q d m A^q + (2 \eta)^q \sum_{j \in J_b} (p_{j}(k))^q
\end{align*}
Since all machines are identical, the second term of the objective (\ref{eqn:main-obj}) is the same for all feasible assignments. Hence we have shown that our final solution is $(4\eta)^q d$-competitive against the optimal fractional solution w.r.t. objective (\ref{eqn:main-obj}). 
 Since $q \geq \log d$, $(4\eta)^q d = O(\log d / \log \log d)^q$, as desired.

\section{Heterogeneous Machines}
%
	\label{sec:hetero-related}

In this section we give our $\Omega(\log m)$ lower bound for related machines with heterogeneous speeds (the second part of Theorem \ref{thm:related-makespan}) , i.e., the speed vector for a fixed machine need not be uniform. This result also extends to a $\Omega(\log d + q)$ lower bound for generic $q$-norms, thereby showing 
a $\Omega(\log m + \log d)$ lower bound for the makespan case when $q = \log m$. 



We (the adversary) construct our online lower bound instance as follows. Let $d = 2h+ 1$ be the number of dimensions; there will be $2^h$ machines in total. All
speeds will be either be 1 or arbitrarily slow; for simplicity, we will just say these machines have speed 0. 
To define each speed $s_{i}(k)$, we first pair off $2h$ of the total  $2h + 1$ dimensions
into $h$ pairs, and order these pairs $1, \ldots, h$ arbitrarily; we will call the remaining dimension that is not 
paired the {\em aggregate} dimension; we will call the other dimensions that are paired {\em pattern} dimensions.  

For each pair of pattern dimensions $(k,k')$ and a fixed machine $i$, we will define machine speeds so that either 
$s_i(k) =1$ and $s_i(k') = 0$ or vice versa.  We say that
$(k,k')$ has {\em speed pattern} $A$ in the former case and speed pattern $B$ in the latter. 
To define speeds over all machines in pattern dimensions, we can 
think of taking the set of all $2^h$ strings $A$s and $B$s of length $h$, mapping each one to a unique machine, and then using the string to define 
the corresponding speed pattern. For example, if we map string $t$ to machine $i$ and the $\ell$th character of $t$ is $B$, then 
for the $\ell$th dimension pair $(k, k')$ we set $s_i(k) = 0$ and $s_i(k') = 1$. Finally, we will simply fix the speed of all machines in the aggregate 
dimension to be 1. This completes the definition of machine speeds in the instance. 
  
Now we describe the job sequence for the instance. Jobs will be issued in $h$ rounds $1, \ldots, h$, one for each dimension pair. Throughout 
the instance, we maintain a set of {\em active} machines in which the algorithm can still use; in other words, jobs will be defined so that they 
cannot be assigned to inactive machines.  Denote the set of active machines at the beginning of round $\ell$ 
as $T_\ell$. At the start of the instance all machines are active, and then each round, the number of active machines is halved, where the goal is to limit the algorithm to machines 
that have already been heavily loaded in the aggregated dimension. 

The adversary maintains active machines as follows:
 Suppose we are in the $\ell$th round of the instance. For this round, we will call a machine an {\em $A$ machine} if it has speed pattern $A$ in the $\ell$th dimension pair; 
 $B$ machines are defined similarly, and inductively assume there are an equal amount of $A$ and $B$ machines in $T_\ell$. 
  We will issue a set of jobs $J_\ell$ such that $|J_\ell| = |T_\ell| = m/2^{\ell-1}$, i.e., we issue as 
 many jobs as there are active machines. After the algorithm assigns the jobs in $J_\ell$, we then observe which set, the $A$ machines or $B$ machines,
 has received the majority of the load among machines in $T_\ell$ in the aggregate dimension up until this point in the instance.
 We will then define future jobs so that they are limited to this more heavily loaded set of machines. For example, 
letting $(k, k')$ denote the $\ell$th dimension pair, if machines in $T_\ell$ with pattern $A$ have received a majority of the jobs up until this point, then for all future jobs $j$ after this round
we define $p_j(k) = 1$ and $p_j(k') = 0$ so that the algorithm is forced to continue to use these machines. We will call this the {\em majority speed pattern} 
for round $\ell$.  We will also define each job so that 
it has load 1 in the aggregate dimension, and the loads for dimension pairs $\ell + 1, \ldots, h$ are defined to be 0.
 This completes the description of the construction, 
and one can verify that this induction is well defined. 

The resulting instance will force a makespan of $h = \Omega(\log m)$ on some machine in the aggregate dimension. This claim is implied by the following 
lemma:
\begin{lemma}
\label{l:makespanAlgo}
The average load on active machines in the aggregate dimension at the start of round $\ell+1$ is at least $\ell$. 
\end{lemma}

\begin{proof}
Consider the start of round $\ell$, and inductively assume the average load on active machines $T_\ell$ is at least $\ell-1$.  
Recall that the number of active machines $|T_\ell| = m/2^{\ell-1}$ at the beginning of this round. Since we issue $m/2^{\ell-1}$ jobs and they can only go to 
active machines, the average load for $T_\ell$ machines increases by 1, i.e., it is now at least $\ell$. Furthermore, since we pick the majority speed pattern based on 
which pattern currently has more load in the aggregate dimension, it is not hard to verify that the average for these $m/2^{\ell}$ machines must also be at least $\ell$. Since 
these machines with the majority speed pattern will be the new active machines for round $\ell+1$, the 
proof of the lemma now follows by induction. 
\end{proof}

To complete the argument, observe that it is possible to ``reverse" the decisions of the algorithm to get a makespan of at most 2 on all machines and dimensions. In particular, the optimal solution assigns all jobs in the $\ell$th round to the machines that do not correspond to the majority speed pattern in the $\ell$th dimension pair (i.e., if the majority speed pattern was $A$ for a round, 
then all jobs are assigned to $B$ machines, and vice versa).  Since in each round half the machines are $A$ and $B$ machines, respectively, 
and we issue as many jobs as there are active machines, 
this will produce a load of 2 on the machines that do not correspond to the majority speed pattern. This completes our proof for the second part of Theorem \ref{thm:related-makespan}.

    We now extend the above lower bound to show a lower bound of $\Omega(\log d + q)$ for case when each dimension can be evaluated with arbitrary $q$-norm  for $1 \leq q \leq \log m$. 
In the above construction, the load vector in the aggregate dimension at the end 
of the instance has load vector identical of that in the $\Omega(q)$ 
lower bound for the single-dimensional unrelated machines lower bound (see \cite{AwerbuchAGKKV95}), and thus the above construction also gives a lower bound of $\Omega(q)$. To obtain a lower bound of $\Omega(\log d)$, we add $m$ {\em additional} dimensions $1, \ldots, m$ to the above construction. Note that now $d = \Theta(m)$. 
The speed in additional dimension $i$ 
is $1$ on machine $i$ and arbitrarily fast on all other machines. These additional dimensions receive the same load that the aggregate does. Based on the construction, 
there will some additional dimension $i'$ with load $\Omega(\log m)$ on machine $i'$ at the end of the instance (the machine that produces a load of $\Omega(\log m)$ in the aggregate dimension). 
Note that the optimal solutions obtains a $q$-norm of $(2^q)^{(1/q)} = O(1)$ on all additional dimensions, whereas the algorithm's solution has 
produced a $q$-norm of $((c \log m)^q)^{(1/q)} = \Omega(\log d)$ for additional dimension $i'$. This completes the extension. 

\section*{Acknowledgement}
We thank Janardhan Kulkarni for many enlightening discussions 
in the early stages of this work.

\bibliographystyle{plain}
\bibliography{ref}

\appendix

\section{Counterexample for Slowest-Fit in Vector Scheduling}
\label{app:example}

The previously known scalar scheduling algorithm~\cite{BermanCK00} for related machines with the makespan norm only loses a constant factor by using {\em slowest-fit}: assign a job to the slowest machine that can accommodate it without exceeding the desired competitive ratio. What if we use the same rule for assigning jobs in vector scheduling for the makespan norm? Unfortunately, this strategy fails.

\smallskip
\noindent
\emph{Example:} Consider a set of homogeneous  related machines where there are $2^g$ machines of speed $1 / 2^{g}$, $g \in \{ 1, 2, \cdots , d\}$ -- let's index this group by $g$. Let $c$  be the desired competitive ratio or equivalently the maximum average load we allow for each group. Note that all groups have an equal `processing power,' $2^g \cdot 1 / 2^{g} = 1$. Any job released is sufficiently small  so that it can be assigned to any group, i.e., $2^d \cdot ||p_j||_\infty \leq 1$ for all $j$. We release jobs in $d$ phases. In the $g$th phase, every job has size $1 / 2^d$ on dimension $g$, and extremely tiny sizes on the other dimensions. 
There will arrive $c \cdot 2^d$ such jobs in this phase. In the spirit of slowest-fit, these jobs will be assigned to group $g$, eventually making the group hit the threshold $c$ on the average load on dimension $g$. Note that other dimensions are barely used. After all the $d$ phases, from $d$ to $1$, we now release tiny jobs with size $1 / 2^d$ on all dimensions. However, every group has hit the predetermined threshold on a distinct dimension, thus can't accept any more jobs. In contrast, it is easy to see an optimal schedule with makespan $c / d$ (ignoring the extremely tiny sizes).

\smallskip
The problem with slowest-fit is that it excessively preserves fast machines for big jobs that may arrive in the future. In particular, it fails to realize in the above instance that all the groups have exactly the same processing power. This suggests that the slowest-fit strategy would work better if we can ensure that the slower groups have larger processing power, and therefore should receive most of the jobs. We artificially ensure this by grouping machines not by speed, but in a way such that the total processing power of the groups increases exponentially as we move to slower machines. While this creates the desired distribution of processing power, we no longer have the property that the machines in the same group have similar speeds. However, we manage to show that we can replace the (actual) machines in each group by a set of (simulated) identical machines with the same cumulative processing power, but with speed equal to that of the slowest machine in the group, without increasing the optimal makespan by more than a constant factor. This constitutes our machine smoothing technique that is given in Section~\ref{sec:grouping}.
\section{Impossibility for All Norms Minimization in Vector Scheduling}
	\label{sec:no-all-norms}

In this section, we provide an instance that rules out all norms minimization even for related machines with homogeneous speeds. This will distinguish related machines from identical machines, for which a logarithmic competitive algorithm was shown for all norms minimization \cite{ImKKP15}. 

\smallskip
\noindent
\emph{Instance.}  There are two type of machines, fast and slow. There are $t$ fast machines with speed 1 and $t^2$ slow machines with speed $1 /  \sqrt t$. The number of dimensions $d = t^2 +1$. There are $t^2$ jobs, and each job has size 1 on a distinct dimension and size $1/t$ on a dimension that is shared by all jobs -- we call this dimension the common dimension; we call the the other dimensions dummy dimensions.

\smallskip

If we place an arbitrary set of $t$ jobs on each fast machine, the makespan is 1, and the $L_4$ norm of the loads is 1 on any dummy dimension and $t^{1/4}$ on  the common dimension. Now let's see how the makespan norm and $L_4$ norm change when we assign each job to a distinct slow machine. Note that the makespan is now $\sqrt t$. The $L_4$ norm also increases to $\sqrt t$ on any dummy dimension, but decreases to $((1 / \sqrt t)^4 t^2)^{1 / 4} = 1$ on the common dimension. Thus one can improve some norm on a specific dimension by a factor polynomial in $d$ while sacrificing others.

\section{Proof of Claim \ref{clm:tied}}

We recall the claim: 
\smallskip
{\em If a job $j$ is assigned in proportion to $|G|\cdot s_G^\gamma$ among machine groups
    $G$ with identical values of $\alpha_{Gj}$, where $\gamma = q/(q-1)$, then the value 
    of $\alpha_{Gj}$ remains equal for these machine groups after the assignment.}
    
\begin{proof}
Recall that 
\begin{equation}
\label{eq:alphadef}
\alpha_{Gj} := q \cdot (\Lambda_G)^{q-1} \cdot\frac{p_j}{s_G}
\end{equation}
Therefore its derivative with respect to an assignment $x_{ij}$ is:
\begin{equation*}
\frac{d \alpha_{Gj}}{d x_{ij}} = q(q-1)\cdot (\Lambda_G)^{q-2} \cdot\frac{p_j^2}{s_G^2}
\end{equation*}
Substituting for $\Lambda_i$ using \eqref{eq:alphadef} we have:
\begin{equation}
\label{eq:sublamb}
\frac{d \alpha_{Gj}}{d x_{ij}} = q(q-1)\cdot \left(\frac{s_G \alpha_{Gj}}{p_j\cdot q}\right)^\frac{ q-2}{q-1} \cdot\frac{p_j^2}{s_G^2}
\end{equation}
To keep $\alpha_{Gj}$ values equal while dividing  $x_{ij}$ infinitesimally among the groups, we 
should assign mass inversely proportional to $\frac{d \alpha_{Gj}}{d x_{ij}}$ times 
$|G|$ to each group $G$. However, since all $G$ already have equal $\alpha_{Gj}$
upon the assignment, all terms in $\frac{d \alpha_{Gj}}{d x_{ij}}$ except 
for $S_G$ are common across these groups. Thus, each group should receive 
mass in proportion to $S_G^{2-(q-2)/(q-1)} |G| = S_G^\gamma|G|$. 

\end{proof}












\end{document}